\documentclass[sigconf]{acmart}
\AtBeginDocument{%
  \providecommand\BibTeX{{%
    \normalfont B\kern-0.5em{\scshape i\kern-0.25em b}\kern-0.8em\TeX}}}

\usepackage{amsmath,amsfonts}
\usepackage{algorithmic}
\usepackage{graphicx}
\usepackage{listings}
\usepackage{makecell}
\usepackage{multirow}
\usepackage{graphicx}
\usepackage{svg}
\usepackage{xspace}
\usepackage{dsfont}
\usepackage{lipsum}
\usepackage{subfigure}
\usepackage{centernot}
\usepackage{diagbox}
\usepackage{tabularx}

\newcolumntype{P}[1]{>{\centering\arraybackslash}p{#1}}
\usepackage[ruled,vlined,linesnumbered]{algorithm2e}

\SetCommentSty{mycommfont}
\usepackage{pifont}
\usepackage{enumitem}

\usepackage{caption}

\usepackage{xcolor}
\usepackage{xcolor,colortbl}
\definecolor{lightgray}{gray}{0.93}
\definecolor{slightgray}{gray}{0.98}
\definecolor{darkgray}{gray}{0.77}

\definecolor{amber}{rgb}{1.0, 0.49, 0.0}

\newcommand{\sidong}[1]{\textcolor{amber}{#1}}

\newcommand{\xmark}{\ding{55}}
\newcommand{\cmark}{\ding{51}}
\newcommand{\tool}{{MUD}\xspace}

\usepackage{framed}
\definecolor{formalshade}{rgb}{0.95, 0.95, 1}
\definecolor{mygray}{gray}{0.4}
\definecolor{lightgray}{gray}{0.93}

\copyrightyear{2024}
\acmYear{2024}
\setcopyright{rightsretained}
\acmConference[CHI '24]{Proceedings of the CHI Conference on Human Factors in Computing Systems}{May 11--16, 2024}{Honolulu, HI, USA}
\acmBooktitle{Proceedings of the CHI Conference on Human Factors in Computing Systems (CHI '24), May 11--16, 2024, Honolulu, HI, USA}
\acmDOI{10.1145/3613904.3642350}
\acmISBN{979-8-4007-0330-0/24/05}

\begin{document}

\title[Towards a Large-Scale and Noise-Filtered UI Dataset for Modern Style UI Modeling]{MUD: Towards a Large-Scale and Noise-Filtered UI Dataset for Modern Style UI Modeling}


\author{Sidong Feng}
\affiliation{%
  \institution{Monash University}
  \city{Melbourne}
  \country{Australia}}
\email{sidong.feng@monash.edu}

\author{Suyu Ma}
\affiliation{%
  \institution{Monash University}
  \city{Melbourne}
  \country{Australia}}
\email{suyu.ma1@monash.edu}

\author{Han Wang}
\affiliation{%
  \institution{Monash University}
  \city{Melbourne}
  \country{Australia}}
\email{han.wang@monash.edu}

\author{David Kong}
\affiliation{%
  \institution{Monash University}
  \city{Melbourne}
  \country{Australia}}
\email{dkon0020@student.monash.edu}

\author{Chunyang Chen}
\affiliation{%
  \institution{Technical University of Munich}
  \city{Heilbronn}
  \country{Germany}}
\email{chun-yang.chen@tum.de}

\renewcommand{\shortauthors}{Feng et al.}

\begin{abstract}
The importance of computational modeling of mobile user interfaces (UIs) is undeniable. However, these require a high-quality UI dataset. Existing datasets are often outdated, collected years ago, and are frequently noisy with mismatches in their visual representation. This presents challenges in modeling UI understanding in the wild. This paper introduces a novel approach to automatically mine UI data from Android apps, leveraging Large Language Models (LLMs) to mimic human-like exploration. To ensure dataset quality, we employ the best practices in UI noise filtering and incorporate human annotation as a final validation step. Our results demonstrate the effectiveness of LLMs-enhanced app exploration in mining more meaningful UIs, resulting in a large dataset \tool of 18k human-annotated UIs from 3.3k apps. We highlight the usefulness of \tool in two common UI modeling tasks: element detection and UI retrieval, showcasing its potential to establish a foundation for future research into high-quality, modern UIs.
\end{abstract}


\maketitle

\section{Introduction}
As mobile apps increasingly become integral to daily life, research interest in developing various applications based on mobile UI screens has surged. 
These applications include UI element detection~\cite{chen2019gallery}, screen embedding~\cite{li2021screen2vec}, widget captioning~\cite{li2020widget}, icon labeling~\cite{feng2021auto}, and screen summarization~\cite{wang2021screen2words}, all of which enhance the interactive capabilities and accessibility of mobile phones.
These tasks typically depend on mobile UI data, specifically UI screenshot images, and the view hierarchy, which represent the elements contained on the screen, their attributes (e.g., position, dimensions), and the structural relationships between them.

Towards that goal, several mobile UI datasets have been collected, including ERICA~\cite{deka2016erica}, Gallery DC~\cite{chen2019gallery}, Guigle~\cite{bernal2019guigle}, VINS~\cite{bunian2021vins}, AMP~\cite{zhang2021screen}, and Swire~\cite{huang2019swire}.
However, these existing datasets are either inaccessible or insufficient for data-driven modeling purposes.
Rico~\cite{deka2017rico} is the largest publicly available mobile UI dataset, comprising 66k UIs from 9.7k Android apps.
It has been a primary data source for much UI understanding research and has been expanded to support numerous downstream applications.
However, Rico's dataset, collected in early 2017, has not been updated since.
We conducted a small empirical study comparing the UIs in the Rico dataset with the latest UIs. 
The latter is designed in a modern style, boasting appealing visual appearances, user-friendly visual hierarchies, simplified user interactions, and readable typography. 
As a result, models trained on the outdated UIs in the Rico dataset may exhibit degraded performance on newer apps with updated design aesthetics.

On the other hand, many studies have identified that view hierarchies are often noisy~\cite{li2020mapping}.
A small-scale pilot study on 500 UIs in the Rico dataset from our study in Section~\ref{sec:noise} reveals three major issues with these view hierarchies related to the UI screens.
First, UIs are typically captured at runtime, but UI rendering may take time, leading to the capture of partially rendered UIs.
Second, due to the nature of UI framing, developers may use views to cover previous elements for simplicity, resulting in overlaid view hierarchies.
Third, there remains a significant percentage of duplicate UIs in the dataset. 
In fact, these problematic issues are not beneficial as either input signals or output labels and might even negatively impact UI modeling performance. 
While many studies~\cite{bunian2021vins, li2020mapping} propose noise removal in Rico, this leaves the dataset on a smaller scale for less generality of data-driven UI modeling.

In this paper, we present \tool, a large-scale, high-quality mobile UI dataset collected from the most recent apps. 
\tool is constructed by mining Android apps at runtime using a novel automated app exploration approach. 
Drawing inspiration from the success of Large Language Models (LLMs) in conversational chatting as a professional expert, we frame app exploration as a question-and-answer (Q\&A) task, i.e., asking the LLMs to play a role as an app expert to interact with the target app.
Specifically, we provide the LLMs with the context of the current UI information via the view hierarchy and prompt potential interactions to operate on the screen, thereby automatically exploring the apps.
During this automated app exploration, we collect a large number of UIs, many of which could be noisy. 
To address this, we adopt mature techniques with best practices to automatically remove the noisy data in advance.
We then enlist human annotators as the final line of defense to audit the UIs and view hierarchies, ensuring the quality of our dataset.

Our \tool dataset comprises 18,132 unique UIs from 3.3k apps, spanning 33 app categories.
Results indicate that our proposed LLM-enhanced approach can boost 17\% in exploring apps, compared to three state-of-the-art tools.
Additionally, we qualitatively investigate the capabilities of LLMs in app exploration and reveal three key findings in our discussion: semantic text input, compound action, and language insensitivity. 
To evaluate \tool's utility, we apply it to the two most common tasks in literature: UI element detection and UI retrieval. 
Preliminary results demonstrate its value in UI modeling, potentially paving the way for further improvements toward UI intelligence.

To summarize, our paper makes the following contributions:

\begin{itemize}
    \item We carry out an empirical study to examine the data issues present in the widely-used Rico dataset. This investigation motivates us to collect a high-quality dataset in a more modern style for data-driven UI modeling.
    \item We introduce a novel approach that elicits the capabilities of LLMs to automatically mine UIs from apps in a manner that mimics human exploration.
    \item We collect, annotate, and open-source the \tool dataset\footnote{\url{https://github.com/sidongfeng/MUD}}, which comprises 18k unique UI screens, each accompanied by a high-quality view hierarchy. We demonstrate the usefulness of the \tool dataset through two applications drawn from the literature: element detection and UI retrieval. These applications highlight \tool's potential as a valuable data source for extensive UI modeling research.
\end{itemize}

\section{Related Work}
\label{sec:related}
We conduct a review of research in three primary areas related to \tool, including automated app exploration, datasets for UI modeling, and applications of UI datasets.

\subsection{Automated App Exploration}
Previous studies primarily relied on human exploration to mine UIs, a process that can be time-consuming and prohibitively expensive~\cite{bunian2021vins,sahami2013insights,deka2017rico}.
In the same vein, many software testing researchers have developed tools to automatically explore apps to detect bugs and generate test scripts. 
While the target differs, these tools can facilitate our task, reducing human effort in UI collection.
One of the earliest initiatives is Monkey~\cite{web:monkey}, Google's official automated app exploration tool, designed to generate random user actions such as clicks, touches, or gestures, as well as several system-level actions on the UI.
Subsequent efforts have focused on improving randomness strategies~\cite{mao2016sapienz,machiry2013dynodroid}. 
However, the random-based testing strategy often fails to formulate a reasonable exploration path based on the characteristics of the app, resulting in low exploration coverage and excessive time consumption.

Recent tools~\cite{liu2022guided,li2017droidbot,amalfitano2012using} have leveraged dynamic and static analysis to reverse engineer a stochastic model from UI for more robust automated app exploration. 
Gu et al.~\cite{gu2019practical} introduced a UI event-refinement model that uses UI runtime information to evolve an initial model, generating precise actions.
Su et al.~\cite{su2017guided} proposed Stoat, which assigns different probabilities to UI runtime elements for selection, achieving effective app exploration. 
Degott et al.~\cite{degott2019learning} adopted reinforcement learning to identify valid interactions for a UI element (e.g., a button can be clicked but not dragged) to guide exploration.
Although model-based automated tools can improve exploration coverage, the coverage remains low as these tools do not take human behavior into account.

Researchers have further proposed human-like strategies and designed learning-based automated app exploration tools. 
Zhou et al.~\cite{zhou2021large} introduced a deep learning model that predicts the next element a user may click, based on the user's click history, the structural information of the UI screen, and the current context, such as the time of day.
However, this user information can be challenging to obtain in practice. 
Li et al.~\cite{li2019humanoid} introduced Humanoid, which uses a sequence of UIs captured from actions to learn a model that predicts human-like interactions on the app. 
Nevertheless, it still struggles to fully understand the semantic information of the UI and plan actions according to the dynamic situation of the app.

Following the success of LLMs in many natural language processing tasks, researchers have started exploring their potential for semantic understanding of UIs.
Wang et al.~\cite{wang2023enabling} examined the use of LLMs in enabling more natural and intuitive conversations between users and UIs, such as fact-based question-and-answer (i.e., ``what is the app version number on the UI'').
Feng et al.~\cite{feng2023prompting} introduced innovative prompting techniques to adapt LLMs for replicating bug steps from bug reports on mobile UIs. 
Our study, however, employs the knowledge nested in LLMs to deduce potential human-like actions on mobile UIs for app exploration.
These models are trained on an ultra-large-scale corpus, which includes tutorials or reports on the web with natural-language descriptions of how specific actions can lead to certain outcomes in software.
By eliciting in-context learning from these models, it potentially aids in exploring the apps more effectively.
More recently, Liu et al.~\cite{liu2023fill} proposed QTypist, a fine-tuned LLMs designed to generate text inputs for exploring more UI states in apps.
Unlike exclusive focus on text input generation, our research provides a nuanced, systematic understanding of the capabilities offered by LLMs, including complex compound actions and multilingual comprehension, as discussed in Section~\ref{sec:exploration_coverage}.
It makes this work more generalized for mobile app exploration in a variety of contexts.




\begin{table}
	\centering
	\caption{A comparison of \tool with other \textbf{open-source} UI datasets.}
	\label{tab:relatedwork}
	\begin{tabular}{l|c|c|c|c|c} 
	    \hline
	    \bf{Dataset} & \bf{Year} & \bf{\# Apps} & \bf{\# UIs} & \bf{Mining} & \bf{Cleaned} \\
	    \hline
	    ERICA~\cite{deka2016erica} & 2016 & 2k & 18k & Hybrid & \xmark \\
	    Rico~\cite{deka2017rico} & 2017 & 9k & 66k & Hybrid & \xmark \\
	    VINS~\cite{bunian2021vins} & 2021 & - & 4k & Manual & \cmark \\
	    \tool & 2023 & 3k & 18k & Auto & \cmark \\
	    \hline
	\end{tabular}
\end{table}

\subsection{Datasets for UI Modeling}
There have been several datasets gathered to support web UI modeling.
For instance, Webzeitgeist~\cite{kumar2013webzeitgeist} utilized an automated crawler to mine 103k designs of web pages and associated web elements with extracted properties such as HTML tag, size, font, and color.
WebUI~\cite{wu2023webui} crawled 400k web pages with semantic information to enhance UI understanding of other platforms with web semantics.
Given the portability and convenience of mobile phones in our daily lives, many researchers have shifted their focus to collecting mobile UI datasets.


Several large-scale mobile UI datasets have been introduced for mobile UI modeling. 
Table~\ref{tab:relatedwork} summarizes the key differences between these existing datasets. 
Shirazi et al.~\cite{sahami2013insights} presented the first mobile UI dataset, manually mining 29k UIs from 400 Android apps.
The ERICA dataset~\cite{deka2016erica} provided a user-friendly web interface that allows users to interact with apps installed on Android devices, collecting 18.6k UIs from 2.4k Android apps. 
The Gallery D.C. dataset~\cite{feng2022gallery}, created for UI element detection, was compiled from 5k app introduction UI screenshots from 68k Android apps in the Google Play Store.
The Guigle dataset~\cite{bernal2019guigle} comprised a large corpus of 5k apps with 12k UI screenshots to facilitate design search.
The AMP dataset~\cite{zhang2021screen} collected 77k UI screens from 4k iOS apps.
However, these datasets are either not publicly available or insufficient for data-driven UI modeling.

The Rico dataset~\cite{deka2017rico} is recognized as the largest publicly available mobile UI dataset, containing 66k UI screens from 9.7k Android apps. 
It has been the primary data source for numerous UI modeling research studies~\cite{wang2021screen2words,li2021screen2vec,li2020mapping,feng2023designing}.
Nevertheless, Rico has several shortcomings. 
First, various studies have identified errors and noise in the Rico dataset, such as instances where nodes in the view hierarchy do not align with the screenshot.
To this end, efforts have been made to repair and filter these examples. 
The Enrico dataset~\cite{leiva2020enrico} initially randomly sampled 10k examples from Rico, then cleaned and provided additional annotations for 1,460 of them.
The VINS dataset~\cite{bunian2021vins} manually labeled a highly accurate view hierarchy for 4k UIs from Rico.
The Clay dataset~\cite{li2022learning} aimed to denoise Rico through a pipeline of automated machine-learning models and human annotators to provide accurate element labels.
However, after cleaning, these datasets may not be suitable for data-driven modeling due to their small size. 
Second, Rico was collected in early 2017 and has not been updated since. 
Many of the latest popular apps adhere to newer design guidelines, like Material Design for Android~\cite{web:material}.
Modeling based on an outdated UI dataset may result in decreased performance on new designs, thus limiting the effectiveness of modeling UI understanding in the current apps.

In our study, we construct \tool, a large-scale, high-quality mobile UI dataset, collected from the most recent apps.
We design a novel tool that automatically navigates through these apps, mining a large number of UIs in the process. 
Recent extensive studies~\cite{fok2022large,feng2023efficiency} reveal that even the most recent apps can still harbor noise and errors.
To ensure the dataset's quality, we utilize mature techniques to carefully filter out noise and errors in advance. 
We further enlist human effort to validate the dataset's quality. 
As a result, \tool contains 18k unique UIs from 3k recent popular Android apps. 
We publicly release \tool to encourage further research on modeling UIs in a modern design style.

\subsection{Applications of UI Datasets}


Originally, many studies relied on pixel-based or heuristic matching~\cite{dixon2010prefab,yeh2009sikuli}.
The advent of UI datasets, like those mentioned earlier, has opened up avenues for developing more robust computational models, particularly those derived from visual data.
In this paper, our focus is on two prevalent uses of UI datasets, namely: (i) element detection at the element level, and (ii) UI retrieval at the screen level.

Element detection is a process that pinpoints the location and type of UI elements within a screenshot. 
This process forms the basis for many subsequent tasks, such as the repair of accessibility metadata~\cite{zhang2021screen,chen2023unveiling,feng2023video2action}, software testing~\cite{xie2020uied,chen2020object,xie2022psychologically,feng2023read,feng2023towards}, and code generation~\cite{chen2018ui,moran2018machine}.
Zhang et al.~\cite{zhang2021screen} suggested an on-device method for element detection on the screen to support accessibility features by training a Faster R-CNN model.
However, the robustness and generality of these models are significantly impacted by the noise present in the UI datasets~\cite{bunian2021vins}.
We discovered that our large-scale clean UI dataset, \tool, could potentially enhance performance as discussed in Section~\ref{sec:downstream_1}.

Other research involves modeling UI screen retrieval to aid designers in data-driven searches, creating design examples by stimulating inspiration, generating new ideas, and examining design decisions from UI repositories. 
Numerous attempts have been made to enhance retrieval algorithms to obtain more relevant UI designs, utilizing deep learning~\cite{chen2020lost,mozaffari2022ganspiration,huang2019swire,feng2021auto, feng2022auto}, reverse engineering~\cite{chen2019gallery,moran2018machine,feng2022gifdroid,feng2022gifdroid1}, and app information~\cite{li2021screen2vec}. 
However, the significance of UI repositories has often been neglected. 
Retrieving UI designs from outdated UI repositories, such as the Rico dataset in 2017, may not inspire modern designs.
Our \tool dataset is driven by the desire to mine the most recent popular UI designs, thereby offering designers a chance to draw inspiration from new design trends.

\section{Empirical Study}
\label{sec:empirical}

In this section, we conducted a small empirical study to understand new UI designs and identify potential noise in existing datasets, thereby highlighting the need for appropriate dataset support.
We chose the Rico dataset~\cite{deka2017rico} as the subject of our study, given its widespread use as a UI dataset. 
We randomly gathered 500 UIs from 27 app categories to form our experimental dataset.
To gain insights into this dataset, we recruited six students as annotators. These students were recruited via the university's internal slack channel and were compensated at a rate of \$12 USD per hour. 
To ensure accurate annotations, we began with an initial training that included reading instructions related to UI understanding, learning the rules of labeling, and passing an assessment test before the participants could begin the various UI annotation tasks.

\begin{figure*}
	\centering
	\includegraphics[width = 0.99\linewidth]{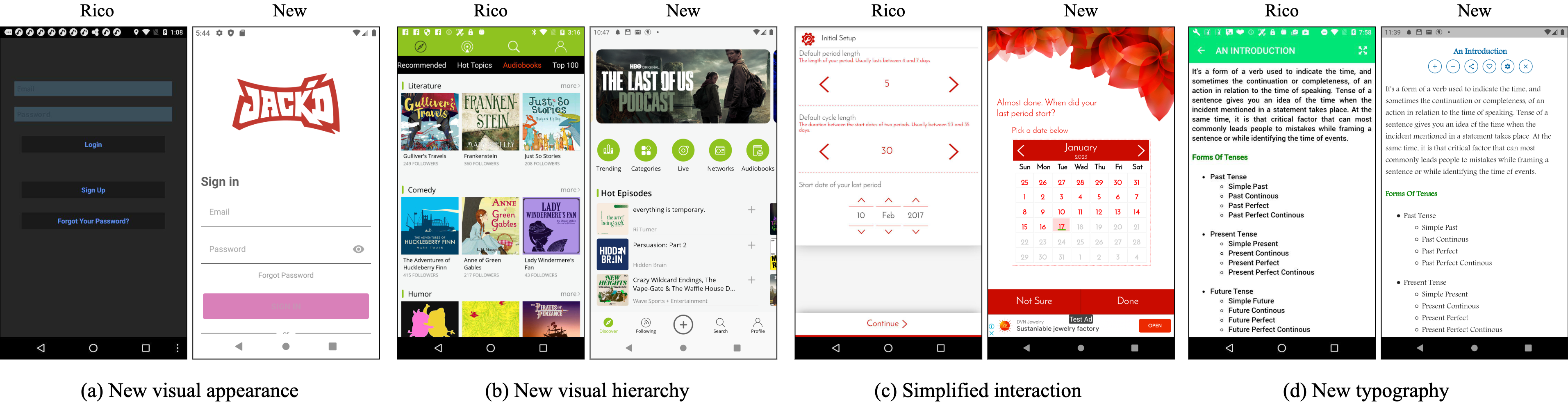}
	\caption{Examples of four types of updates in the new UI design, including visual appearance, visual hierarchy, user interaction, and typography.}
	\label{fig:new}
\end{figure*}

\subsection{Do the apps feature updated UI designs?}
To determine whether the UIs in the Rico dataset have been updated, we compared the UIs to the corresponding latest versions of the apps. 
To do this, we first developed a web crawler to scrape the most recent versions of the apps in our experimental dataset from the Google Play Store, in total, we gathered 132 apps.
We then asked the annotators to manually explore the apps to identify the corresponding UI designs. 
We opted for human exploration because 1) The previously recorded interaction trace to the UIs in Rico may differ in the new app due to new functionalities and usage scenarios; 
2) Automated app exploration tools might not uncover the corresponding UIs. 
As such, we instructed the annotators to familiarize themselves with the UIs and the previous interaction traces in the apps. 
This contextual information, such as previous interaction traces, could potentially help annotators navigate the apps more easily to find the corresponding UIs. 
After an average of 4 hours of human exploration, we obtained 82 pairs of UI designs from both the Rico dataset and the new app version.

To better comprehend the evolution of design styles from old to new, we instructed the annotators to independently categorize the UI pairs using existing UI/UX design knowledge, as documented in resources like The Design of Everyday Things~\cite{norman2013design} and Mobile Design Pattern Gallery~\cite{neil2014mobile}. 
Following the initial categorization, the annotators convened to discuss discrepancies and establish a set of new categories, continuing until a consensus was reached. 
We identified four primary types of UI design updates from Rico, as illustrated in Fig.~\ref{fig:new}.

First, the visual appearance of mobile UI is evolving, encompassing aspects such as color, contrast, and element design.
For instance, as depicted in Fig.~\ref{fig:new}(a), an appealing visual appearance can guide users, optimize readability and accessibility, and ensure a superior user experience.
Second, improvements in the visual hierarchy are being made to better accommodate human perception within the UI. 
Principles such as the Gestalt theory~\cite{fraher2010gestalt}, derived from recent research in psychology and biological vision, provide guidelines for visual UI design cues like connectivity, similarity, proximity, and continuity. 
These principles demonstrated in Fig.~\ref{fig:new}(b), are widely used to enhance user experience and usability. 
Third, interactions are being optimized to simplify the user experience, enabling users to complete tasks more quickly and efficiently, thereby boosting user engagement. 
For instance, the interaction for selecting a date is simplified in the new design shown in Fig.~\ref{fig:new}(c).
Lastly, the importance of typography, previously underestimated in the early years, is increasingly recognized.
A well-designed typeface and style can effectively highlight words and capture users' attention. 
For example, in Fig.~\ref{fig:new}(d), the text in the UI is updated to the ``Footlight'' font, which is easy to read on small screens due to its well-spaced letters and clear strokes.



\begin{figure}
	\centering
	\includegraphics[width = 0.99\linewidth]{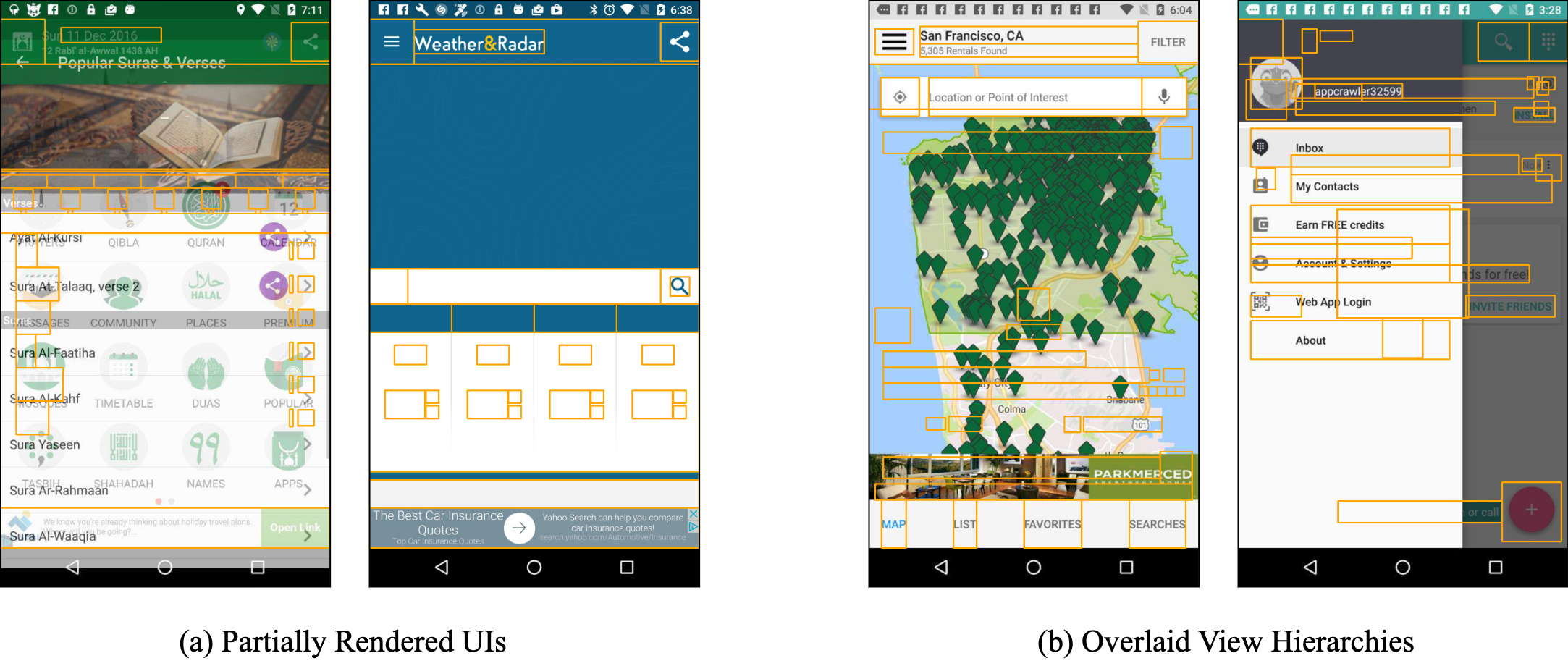}
	\caption{Examples of noises in the Rico dataset. The orange bounding box presents the elements in the view hierarchy.}
	\label{fig:noise}
\end{figure}

\begin{figure*}
	\centering
	\includegraphics[width = 0.75\textwidth]{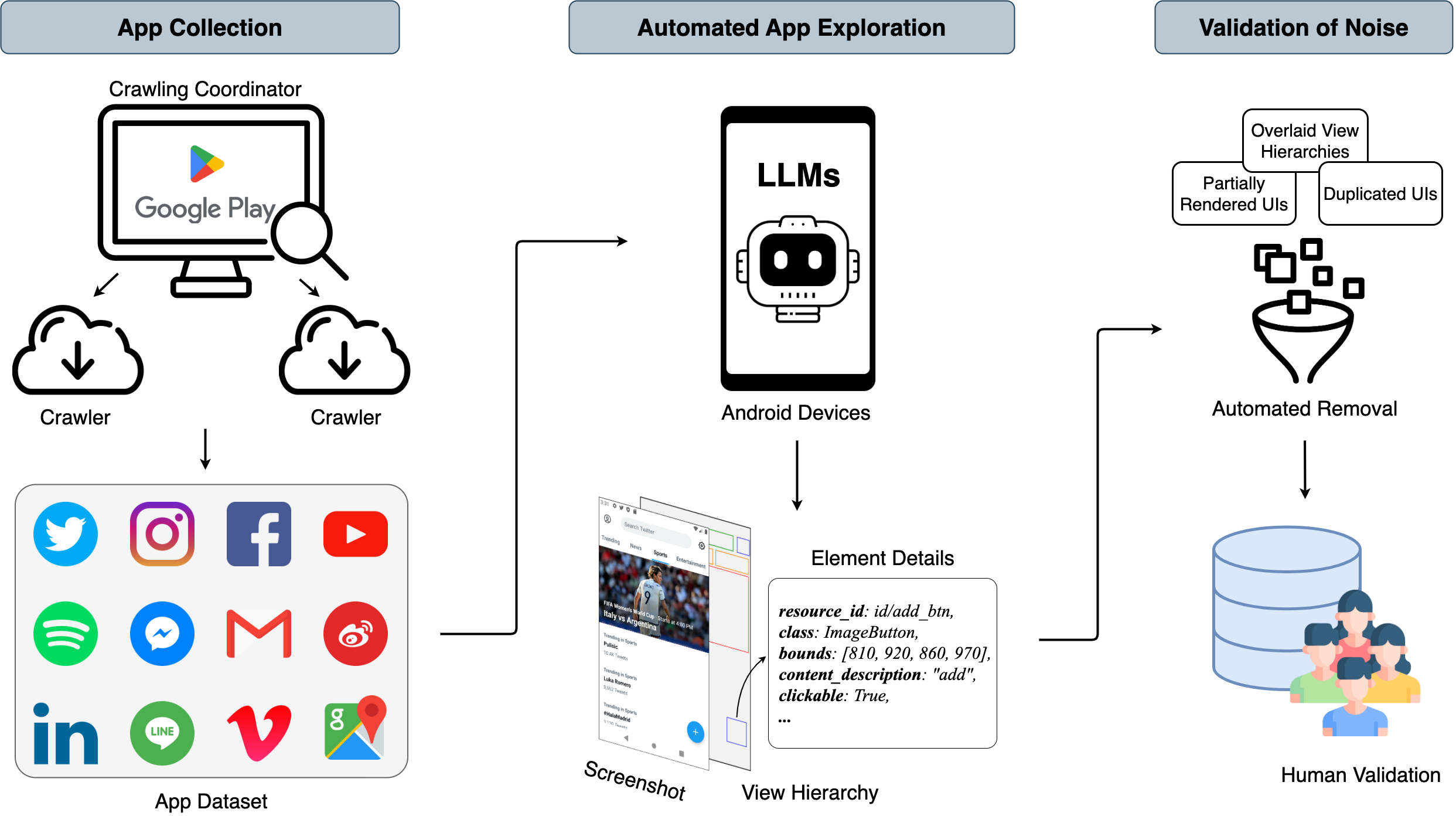}
	\caption{The overview of our dataset collection process.}
	\label{fig:overview}
\end{figure*}

\subsection{What are the prevalent noises in the UI?}
\label{sec:noise}
To examine the extent of noises and errors in the UI datasets, we asked the annotators to label the noisy data in a sample of 500 experimental UIs.
During this manual review process, we observed various types of UI noise. Categorizing these noises could help elucidate the issues and streamline our UI collection process. 
Based on previous studies~\cite{li2022learning,bunian2021vins}, we asked the annotators to independently identify categories and discuss any discrepancies.
Following the Card Sorting method~\cite{spencer2009card}, we identified three prevalent types of noise in the UI datasets:

1) Partially Rendered UIs (16\%). 
UI rendering is the act of generating a frame and displaying it on the screen, including transitioning from the previous page, loading resources from the internet, and drawing UI objects (like buttons) into pixels. 
The duration of this process can vary based on the quality of the app code, device performance, and internet bandwidth. 
However, 16\% of the UIs in our experimental dataset are partially rendered, as shown in Fig.~\ref{fig:noise}(a). This could be due to a short waiting delay during screenshot capturing.

2) Overlaid View Hierarchies (21\%).
Apps often incorporate numerous functionalities, which can lead to overlapped designs in the user interface. 
Some developers might oversimplify app development by overlaying new views on top of the existing UI.
While this doesn't result in any visual differences, as the previous elements are entirely covered by the new view, it does lead to a misalignment in the view hierarchy, as demonstrated in Fig.~\ref{fig:noise}(b).

3) Duplicated UIs (2\%). 
We discovered that 2\% of the UIs are duplicates. 
This could be attributed to two main factors.
First, the process of exploring apps to collect UIs often involves navigating back and forth, which can result in duplicated states. 
Second, certain interactions may trigger a toast~\cite{web:toast}, a brief message displayed on the screen for users, which disappears automatically after a short period of time. 
Consequently, the UIs collected before and after the appearance of the toast are identical, leading to duplication in the dataset.

\vspace{12pt}
\noindent\fbox{
    \parbox{0.98\linewidth}{
        \textbf{Summary}: 
        Upon analyzing 500 UIs from Rico, we found that they are now designed in a more contemporary style. Despite the considerable size of the Rico set, 39\% of its UIs are affected by noise, including partially rendered UIs, overlaid view hierarchies, and duplicated UIs. These old-fashioned UI designs and the presence of noise in Rico could significantly impact data-driven modeling in current UI tasks. This underscores the need for a high-quality dataset featuring UIs designed in the new fashion.
    }
}

\section{The \tool Dataset}
\label{sec:approach}
Given that manually exploring apps can be both time-consuming and labor-intensive, we propose a novel automated app exploration method. 
This method encourages Large Language Models (LLMs) to act as app experts and facilitates interactions within the app.
To address the potential noises outlined in Section~\ref{sec:noise}, we employ mature techniques and best practices from prior works. 
Lastly, as a final safeguard, we implement manual validation to ensure the quality of the dataset.
The overview of our dataset collection is shown in Fig.~\ref{fig:overview}.


\subsection{App Collection}
The quality of apps can directly influence the quality of the UIs.
The more popular the apps are, the higher the quality of the UIs, and the more beneficial they are to data-driven UI modeling. 
To achieve this, we develop a parallelizable, cloud-based web crawler to collect the most popular apps from the Google Play Store. 
Specifically, our crawler is composed of (i) a crawling coordinator server that monitors visited and queued URLs, (ii) a pool of crawler workers that scrape URLs using a headless browser, and (iii) a database service that stores artifacts uploaded by the workers.
The coordinator server distributes seeds to the workers based on the Breadth-First search strategy~\cite{najork2001breadth}, which involves collecting a queue of URLs from a seed list and using the queue as the seed in the second stage of iteration. 
The crawler worker is implemented using a headless framework~\cite{web:chromedriver} to interface with the Chrome browser.
We also introduce simple heuristics to automatically dismiss certain types of pop-ups (e.g., GDPR cookie warnings) to facilitate access to page content. 

\subsection{Automated App Exploration}

\begin{figure*}
	\centering
	\includegraphics[width = 0.85\linewidth]{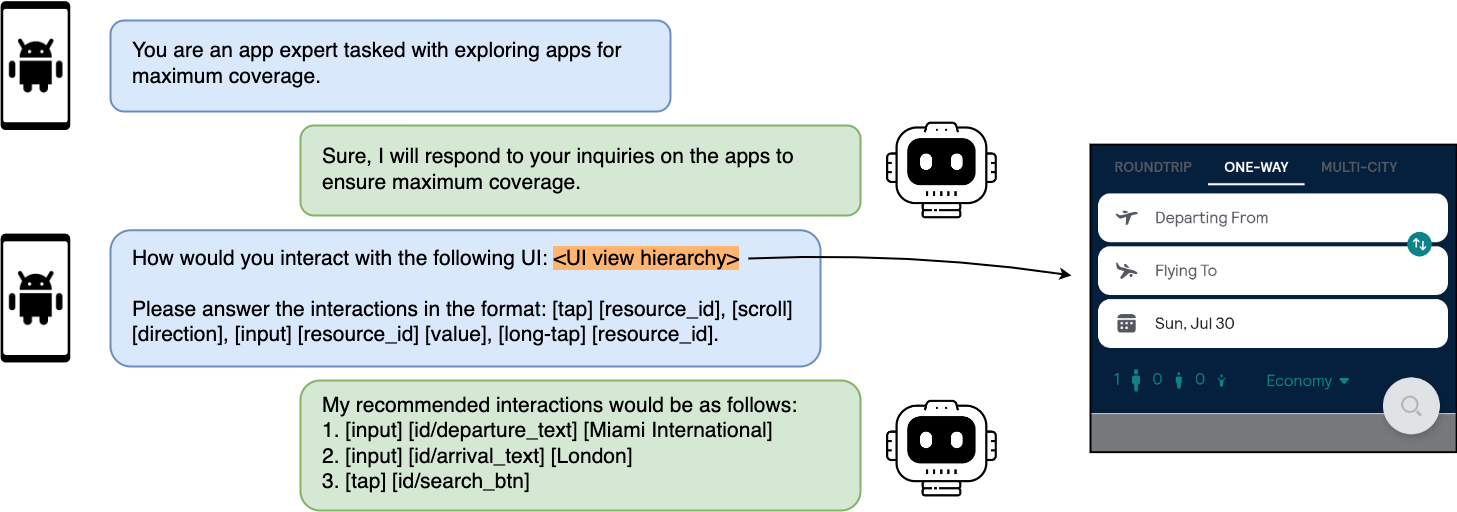}
	\caption{Illustration of our LLMs. We prompt the model to suggest potential interactions, based on the current UI view hierarchy, in order to achieve maximum coverage.}
	\label{fig:llms}
\end{figure*}

Large Language Models (LLMs)~\cite{brown2020language, chowdhery2022palm, liu2019roberta, raffel2020exploring} trained on ultra-large-scale corpora have shown promising performance in areas such as natural language understanding, logical reasoning, and question answering. 
The success of ChatGPT~\cite{web:chatgpt} exemplifies how an LLM can comprehend human knowledge and interact with humans as an informed expert. 
Inspired by ChatGPT, we have reimagined app exploration as a Question \& Answering (Q\&A) task, where we ask the LLMs to act as an app expert and interact with the screen to explore the app.
An overview of our approach is depicted in Fig.~\ref{fig:llms}.
Specifically, we first prompt the LLMs to play the role of app expert by:
\textit{``You are an app expert tasked with exploring apps for maximum coverage.''}
Then, we dump the view hierarchy that represents the current UI screen. 
With the view hierarchy serving as the context for available interactive elements (i.e., those flagged with a \textit{``clickable''} attribute) in the UI, we then prompt the LLMs with the following questions:
\textit{``How would you interact with the following UI <UI>?''}
To further effectively translate the LLMs' responses into actionable operation scripts for executing the app, we also equip the LLMs with a set of interaction primitives, like:
\textit{``Please answer the interactions in the following format: [tap] [resource\_id], [scroll] [direction], [input] [resource\_id] [value], [long-tap] [resource\_id].''}

Our approach is implemented as a fully automated app exploration tool.
For the LLMs, we employ the gpt-3.5-turbo model~\cite{web:chatgpt} from OpenAI.
Since LLMs may generate verbose output (like repeated questions or reasoning), we use \texttt{``[]''} to deduce the specific feedback for operations, such as action, target component resource\_id, and input value. 
Given that LLMs may generate a sequence list of potential actions, i.e., complex compound actions, to facilitate exploration as depicted in Fig.~\ref{fig:llms}, we employ regular expressions to extract the numerical index, serving as the sequence of operations.
We use Android UIAutomator~\cite{web:uiautomator} to dump the UI view hierarchy and Android Debug Bridge (ADB)~\cite{web:adb} to execute the operations on the device.
Additionally, we explore automatic login policies by activating the Google account authentication~\cite{web:login}, which allows a successful login into the Google account to unlock apps and enable the exploration of features hidden behind login screens.
To lessen the noise from partially rendered UIs, as detailed in Section~\ref{sec:noise}, we establish a relatively lengthy waiting period between each operation to allow the UI to fully render. 
According to a small-scale pilot experiment, we have set this waiting time to 2 seconds.

\subsection{Validation of Noise}
\label{sec:approach_noise}
During such automated app exploration, the tool automatically collects pairs of runtime UI screenshots and their corresponding runtime UI view hierarchies. 
Although we have implemented some best practices in the tool to minimize UI noise, such as allowing a long waiting time for partially rendered UIs, some residual noise remains, as discussed in Section~\ref{sec:noise}, such as overlaid view hierarchies and duplicated UIs. 
To address this, we utilize mature techniques from previous studies to mitigate these noises in our UI datasets.

First, we utilize a heuristic method from~\cite{chen2019gallery} to discern duplicated UIs by encoding the underlying element hierarchy of a UI. 
Specifically, we assign a unique type\_index to each type of UI element — for instance, Button is designated as 0, TextView as 1, RatingBar as 2, and so on. 
We then obtain the node\_index (starting at 0) of each UI element in the UI and concatenate an element’s node\_index with its type\_index. 
For example, if the first UI element is a TextView, it can be represented as ``0\_1''. 
Consequently, we generate a representation string for a UI by concatenating the representation strings of all its UI elements. 
We then hash this representation string using the MD5 algorithm~\cite{rivest1992md5} to produce a fixed-length unique identifier for the UI. 
If two UIs share the same identifier, they are considered identical, and the duplicate is removed.

Second, we utilize a deep learning model proposed in~\cite{li2022learning} to automatically identify overlaid view hierarchies. 
Specifically, we frame the detection of overlaid view hierarchies as a binary classification task — determining whether the elements in the view hierarchy are overlaid or misaligned. 
We train a binary classification model using ResNet, with the input being a four-channel matrix. 
The first three channels correspond to the original pixels of the UI image, while the fourth, or mask channel, indicates the bounding box of the element under examination. 
With the mask channel, the model is able to recognize the element's location and concentrate on the element pixels to make a prediction. 
We then filter the overlaid view hierarchy based on the model's prediction of how likely it is that the element is overlaid.

In addition to automated noise removal, we also employ manual validation as the last line of defense. 
We engage the six students mentioned in Section~\ref{sec:empirical} to meticulously validate the dataset.
We follow a similar strategy in ~\cite{li2022learning} to ensure accurate annotations, which creates a web interface to enable participants to annotate the UI dataset efficiently.
The interface (Fig.~\ref{fig:interface} in Appendix) displays a screenshot of the mobile UI, along with the bounding boxes extracted from the view hierarchy. Participants can flag the UI as invalid and select the reasons, such as partially rendered UI, overlaid view hierarchy, duplicate UI, or other reasons.
To further ensure quality, two authors audit the results by randomly sampling 5\% of the labeled UIs during the labeling process for verification.

\subsection{Data Analysis}
\label{sec:data_analysis}
The collection process continued from February 27, 2023, to August 19, 2023, with a collection of 18k unique UIs from 3.3k Android apps. 
In this section, we present a comprehensive analysis of our dataset \tool, including exploration coverage and dataset statistics.

\begin{figure}
	\centering
	\includegraphics[width = 0.98\linewidth]{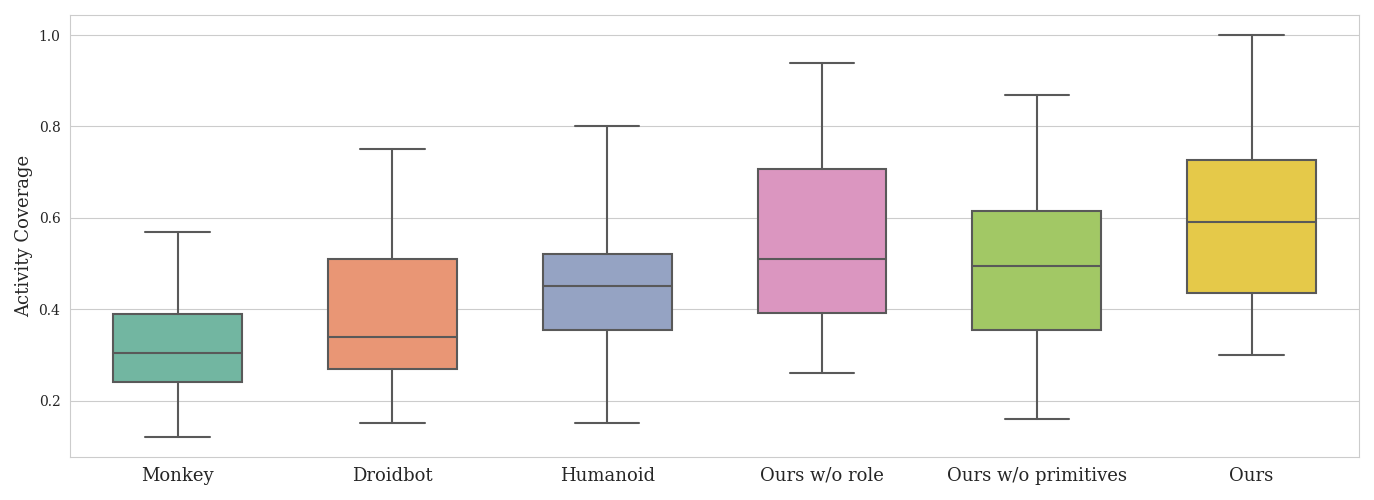}
	\caption{The performance of our LLMs-enhanced exploration method compared to three automated exploration tools, including Monkey, Droidbot, and Humanoid, and two ablation studies, including the prompt without role instantiation and action primitives.}
	\label{fig:coverage}
\end{figure}

\subsubsection{Exploration Coverage}
\label{sec:exploration_coverage}
To measure the coverage benefits of our LLMs-enhanced automated exploration approach, we compare it with three commonly used and state-of-the-art baselines.
There is one random-based tool \textbf{Monkey}~\cite{web:monkey}, one model-based tool \textbf{Droidbot}~\cite{li2017droidbot}, and one learning-based tool \textbf{Humanoid}~\cite{li2019humanoid}.
To further demonstrate the advantage of our approach, we set up two ablation studies.
Given that we propose a role instantiation prompt to define the task objective of app exploration, we consider a variant of our approach without the tailored prompt \textbf{Ours w/o role} to compare the performance of our approach with and without role instantiation.
We further investigate the contribution of action primitives to the prompt, namely \textbf{Ours w/o primitives}, to see the impact of output formatting.
Specifically, we follow the previous work~\cite{li2020mapping} to extract specific action operations from the natural language output.

We use the default configuration settings for each tool and record activity by running the apps for 15 minutes.
We use activity coverage as the evaluation metric, i.e., collecting all the activities defined in each app from AndroidManifest.xml following previous studies~\cite{chen2019storydroid,liu2022guided}, and measuring the percentage of the explored activities during runtime.
However, extracting all the activities from the apps can be time-consuming, especially for closed-source and highly confidential apps.
Therefore, we conduct experiments with 30 apps randomly selected from the dataset, each of which has a rating higher than 4 stars (out of 5) and has been downloaded more than a million times.

Fig.~\ref{fig:coverage} displays the activity coverage of our approach in comparison to the baselines and ablations. 
Our approach achieves a median activity coverage of 0.60 across 30 mobile apps. 
In contrast, Monkey, which relies on random user actions to explore the app, can only achieve an average activity coverage of 0.32. 
Our approach even surpasses the best baseline (Humanoid) in activity coverage by 17\% (0.43 vs. 0.60). 
This demonstrates the effectiveness of our approach in covering the majority of activities within the apps.
We can also see that applying the prompt of action primitives can significantly improve the performance of our approach, leading to an improvement of 11\% in activity coverage.
This is due to the fact that the feedback generated by LLMs tends to be verbose (i.e., repeated questions and reasoning) and ambiguous (i.e., ``create an account name of test'' where ``create'' refers to the ``input'' action), resulting in incorrect action extraction.
In addition, augmenting the prompt with role instantiation leads to a 5\% increase in activity coverage, suggesting that LLMs can focus more effectively on app navigation to trigger additional activities.

\begin{figure}
	\centering
	\subfigure[Valid text input, where the email is entered as ``example@gmail.com'', and the mobile number is a valid 9-digit number.]{
		\includegraphics[width = 0.275\linewidth]{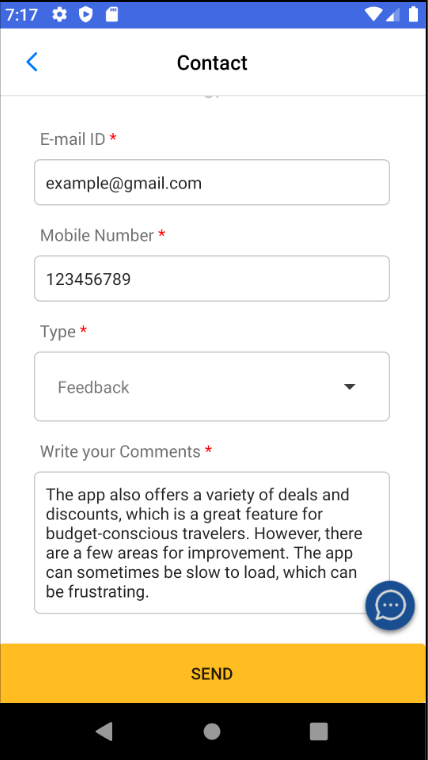}
		\label{fig:finding_1}}
	\hfill
	\subfigure[Compound actions, involve first ticking the checkbox to accept the policy, followed by clicking on the agree button.]{
		\includegraphics[width = 0.275\linewidth]{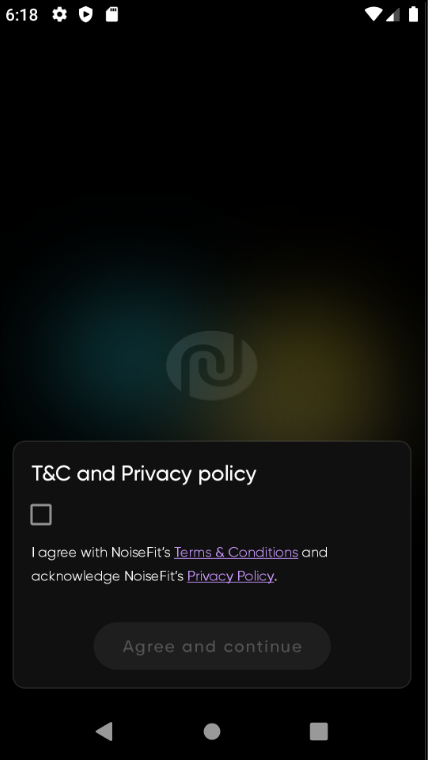}
		\label{fig:finding_2}}
        \hfill
        \subfigure[Multilingual understanding, as it will not be able to proceed to the next UI until the multilingual questions are answered correctly.]{
		\includegraphics[width = 0.275\linewidth]{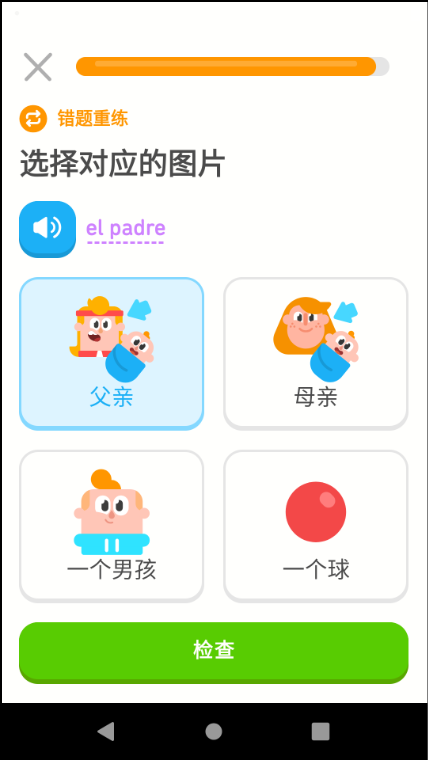}
		\label{fig:finding_3}}
	\caption{Examples of the capability of our approach.}
	\label{fig:finding}
\end{figure}

\begin{figure*}
	\centering
	\includegraphics[width = 0.8\linewidth]{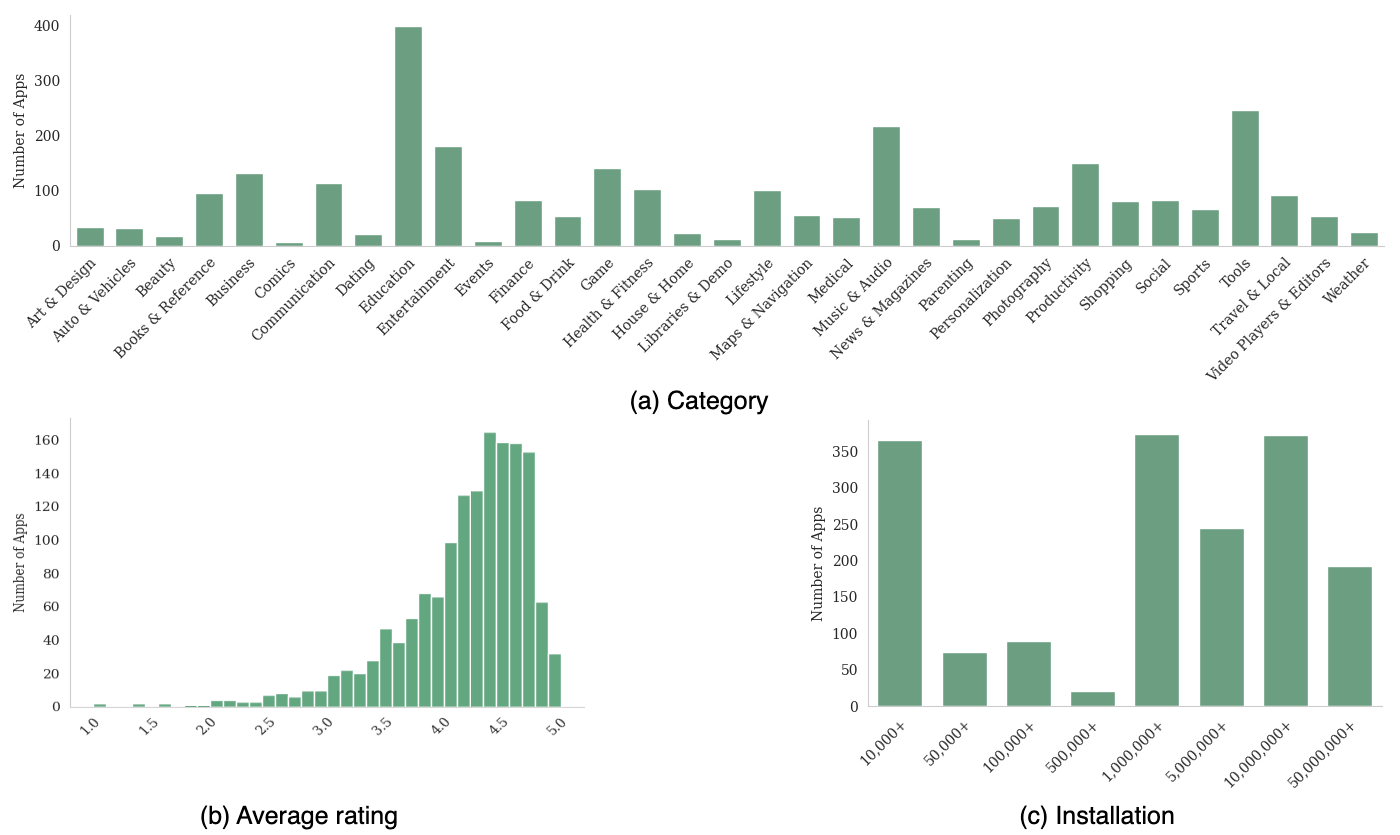}
	\caption{Summary of the app statistics of the \tool dataset: (a) category distribution, (b) average rating distribution, and (c) installation distribution.}
	\label{fig:appstatistic}
\end{figure*}

To fully understand the capability of our approach, we carry out a qualitative study, examining cases where our LLMs-enhanced tool outperforms the baselines.
We identify three key capabilities. 
First, our tool can automatically populate valid text content into the input element, which is crucial for navigating through the page. 
This is primarily due to the LLMs' training on a large-scale corpus, which captures semantically correlated text for input, for example, the email input ``example@gmail.com'' and valid 9-digit mobile number as seen in Fig.~\ref{fig:finding_1}. 
Second, our tool can generate complex compound operations. 
While the baseline focuses on a single action on the UI, sometimes a series of actions in the right order is required to reach the next state. 
This is facilitated by the LLMs' ability to understand the causal relationships between the elements in the UI. 
As illustrated in Fig.~\ref{fig:finding_2}, to get through the state, it first clicks on the checkbox to accept the policy, and then clicks on the agree button. 
Third, our approach is not sensitive to multilingual apps. 
As seen in Fig.~\ref{fig:finding_3}, allows for the exploration of international apps in various languages.
More than that, the app requires that all multilingual questions be answered correctly before it proceeds to the next stage, which our approach has the capability to achieve.
This is because the LLMs are trained on a diverse set of texts in various languages, e.g., at least 95 natural languages for the GPT model~\cite{web:language}.
The capability of our LLM-enhanced approach results in a performance boost of 28\%, 23\%, and 23\% in activity coverage compared to the baseline tools - Monkey, Droidbot, and Humanoid, across 10 (33\%) mobile apps supporting multiple languages included in our experimental dataset.

\subsubsection{Dataset Statistics}
Our dataset is composed of 3.3k apps spread over 33 different categories. 
Fig.~\ref{fig:appstatistic}(a) illustrates the diverse distribution of app categories, covering education, finance, food \& drink, etc.
Fig.~\ref{fig:appstatistic}(b) presents the ratings of these apps, while Fig.~\ref{fig:appstatistic}(c) displays the download statistics of these apps. 
The apps in our dataset have an average download count of 71 million and hold an average rating of 4.17 out of 5.0 stars, emphasizing the popularity and high quality of the apps in our collection.

Our automated app exploration has amassed a collection of 46k UI screenshots, each paired with its corresponding view hierarchy.
We automatically filter out duplicates and overlaid view hierarchies as discussed in Section~\ref{sec:approach_noise}, which results in the removal of 22k UIs (automatically eliminating 85.7\% noises). 
To further ensure the quality of our dataset, we also carry out manual validation, leaving 18k UIs (39.1\%) in the \tool dataset.



\section{Downstream Tasks From \tool}
\label{sec:downstream}
We have showcased the effectiveness of our approach in collecting a high-quality UI dataset, \tool. 
In this section, we further highlight the potential value of \tool across two common UI modeling tasks, namely element detection and UI retrieval.

\begin{table}
	\centering
        \tabcolsep=0.35cm
	\caption{Performance comparison of UI element detection (> 0.5 IoU).}
	\label{tab:element_detection}
	\begin{tabular}{lr|c|c} 
	    \hline
	    \bf{UI Element} & \bf{Count} & \bf{Rico} & \bf{\tool} \\
	    \hline
	    Button & 9,552 & 63.4\% & 75.3\% \\
	    Checkbox & 1,151 & 75.5\% & 84.6\% \\
            EditText & 2,547 & 63.5\% & 75.0\% \\
            Image & 17,843 & 72.4\% & 81.9\% \\
            Radiobutton & 499 & 57.2\% & 68.3\% \\
            Spinner & 255 & 65.3\% & 79.1\% \\
            Switch & 176 & 79.3\% & 91.5\% \\
            Textview & 37,946 & 80.5\% & 90.7\% \\
            Togglebutton & 96 & 88.6\% & 93.7\%\\
	    \hline
            Average & & 71.7\% & \bf{82.2\%} \\
            \hline
	\end{tabular}
\end{table}

\begin{figure*}
	\centering
	\subfigure[]{
		\includegraphics[width = 0.185\linewidth]{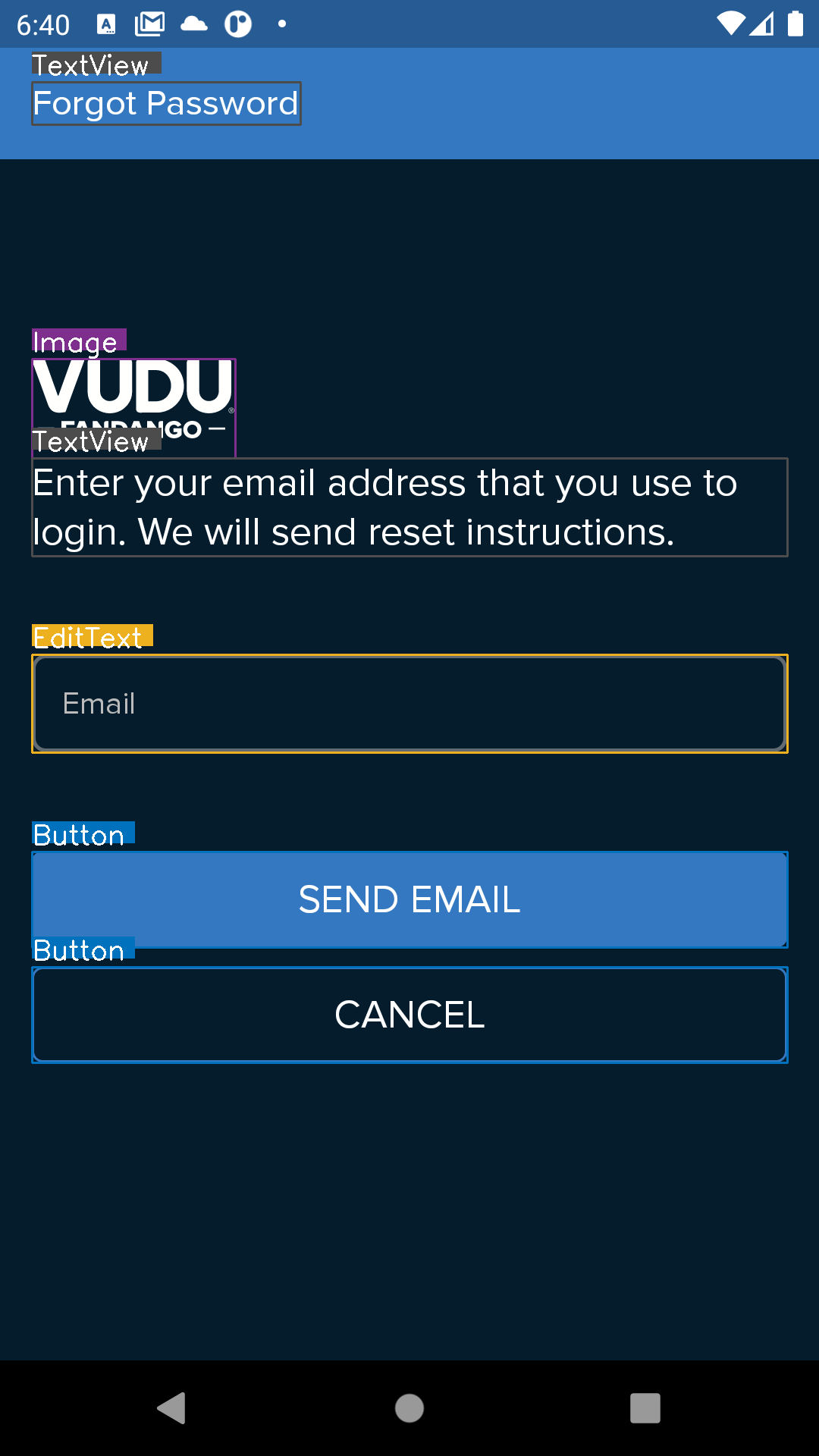}
		\label{fig:vis_121}}
	\hfill
	\subfigure[]{
		\includegraphics[width = 0.185\linewidth]{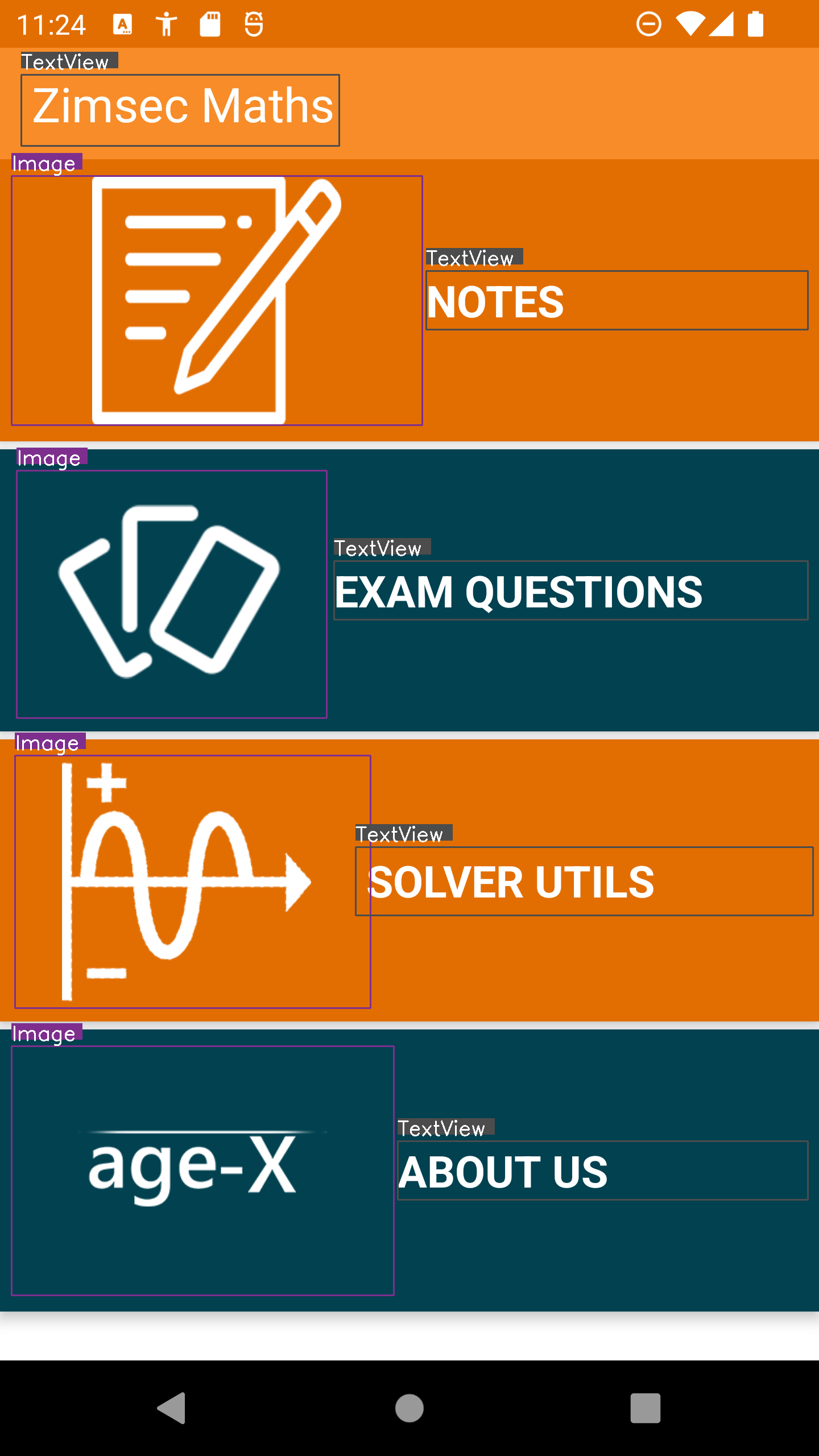}
		\label{fig:vis_243}}
        \hfill
        \subfigure[]{
		\includegraphics[width = 0.185\linewidth]{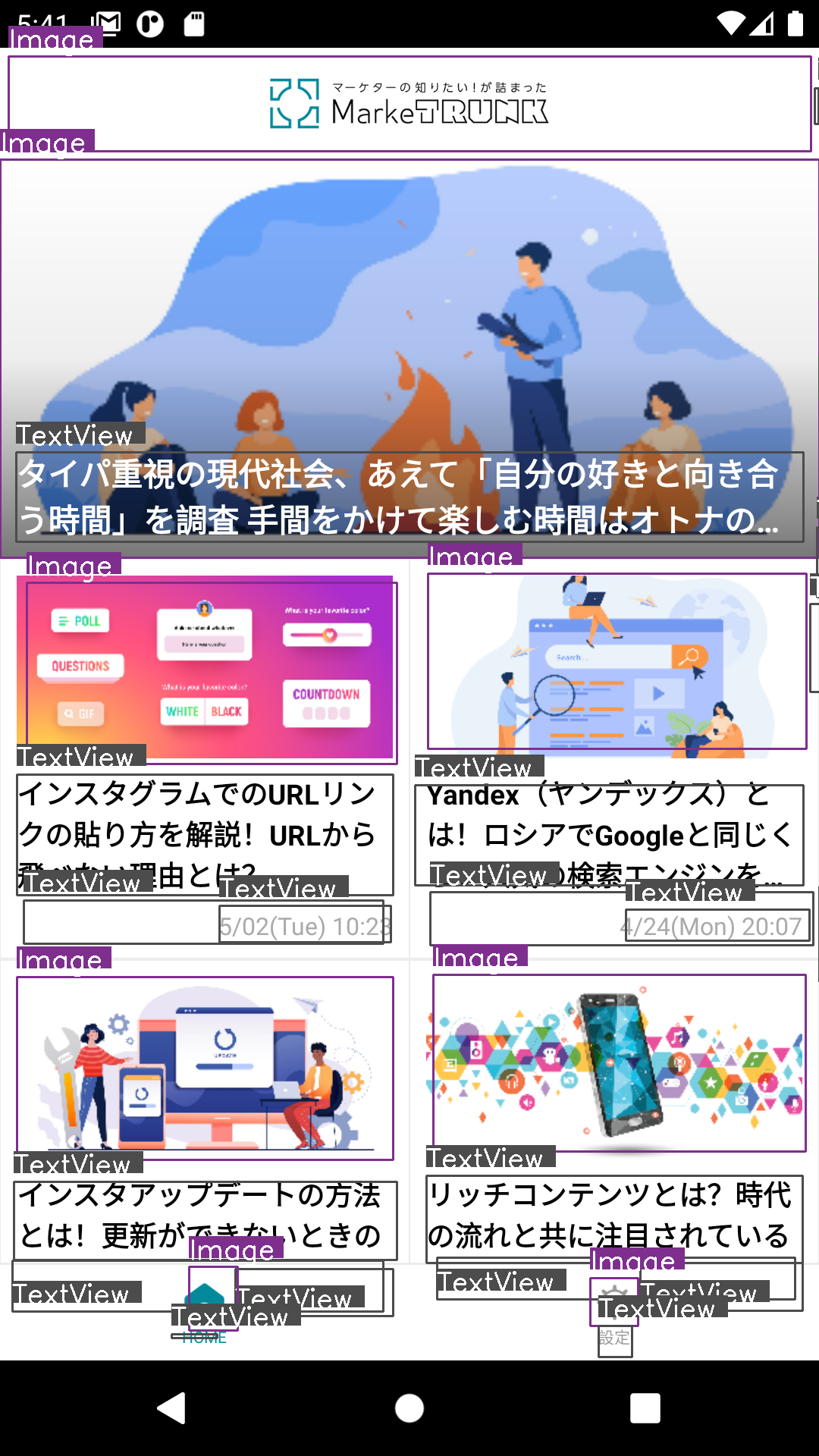}
		\label{fig:vis_607}}
        \hfill
        \subfigure[]{
		\includegraphics[width = 0.185\linewidth]{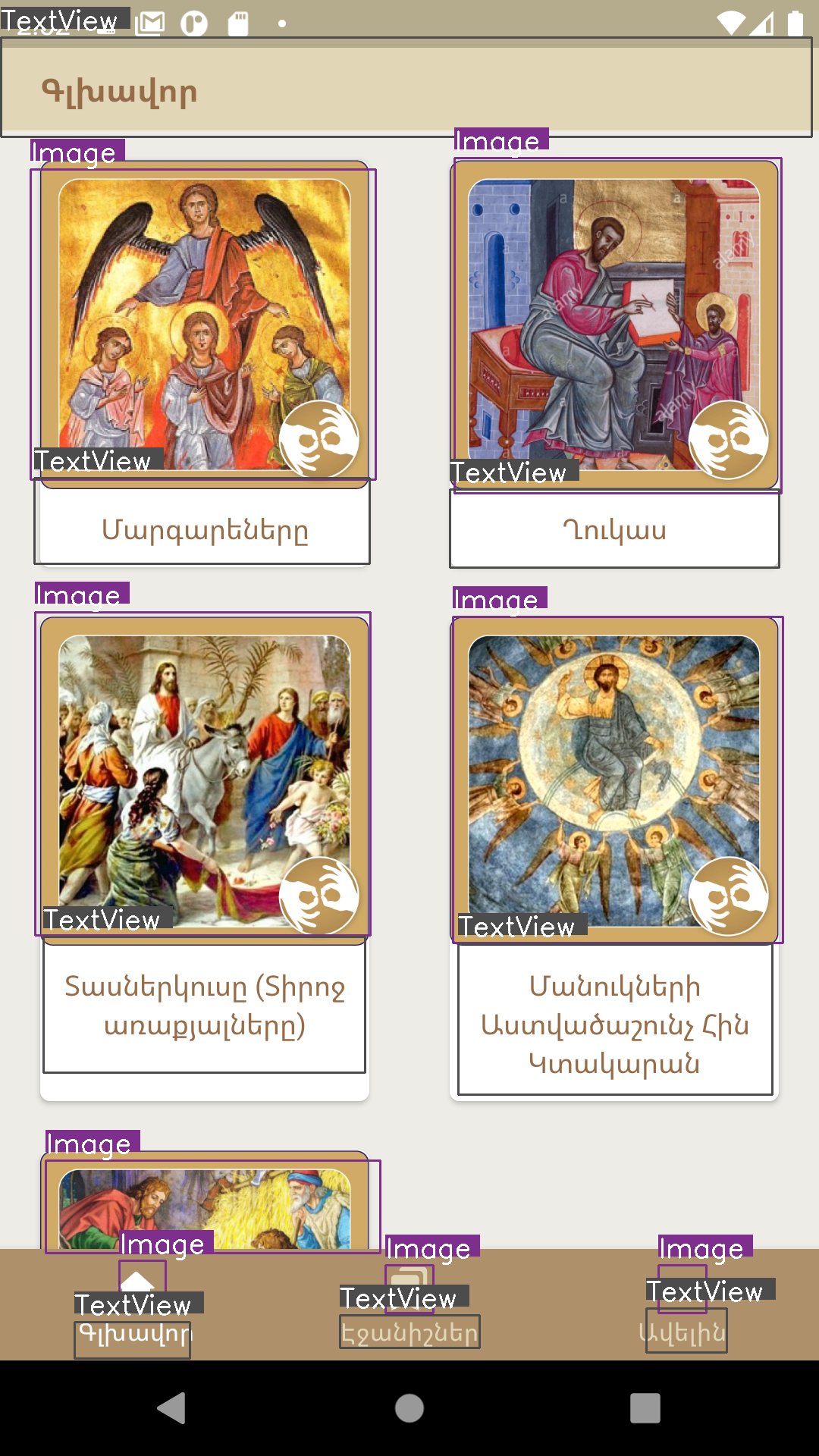}
		\label{fig:vis_7222}}
        \hfill
        \subfigure[]{
		\includegraphics[width = 0.185\linewidth]{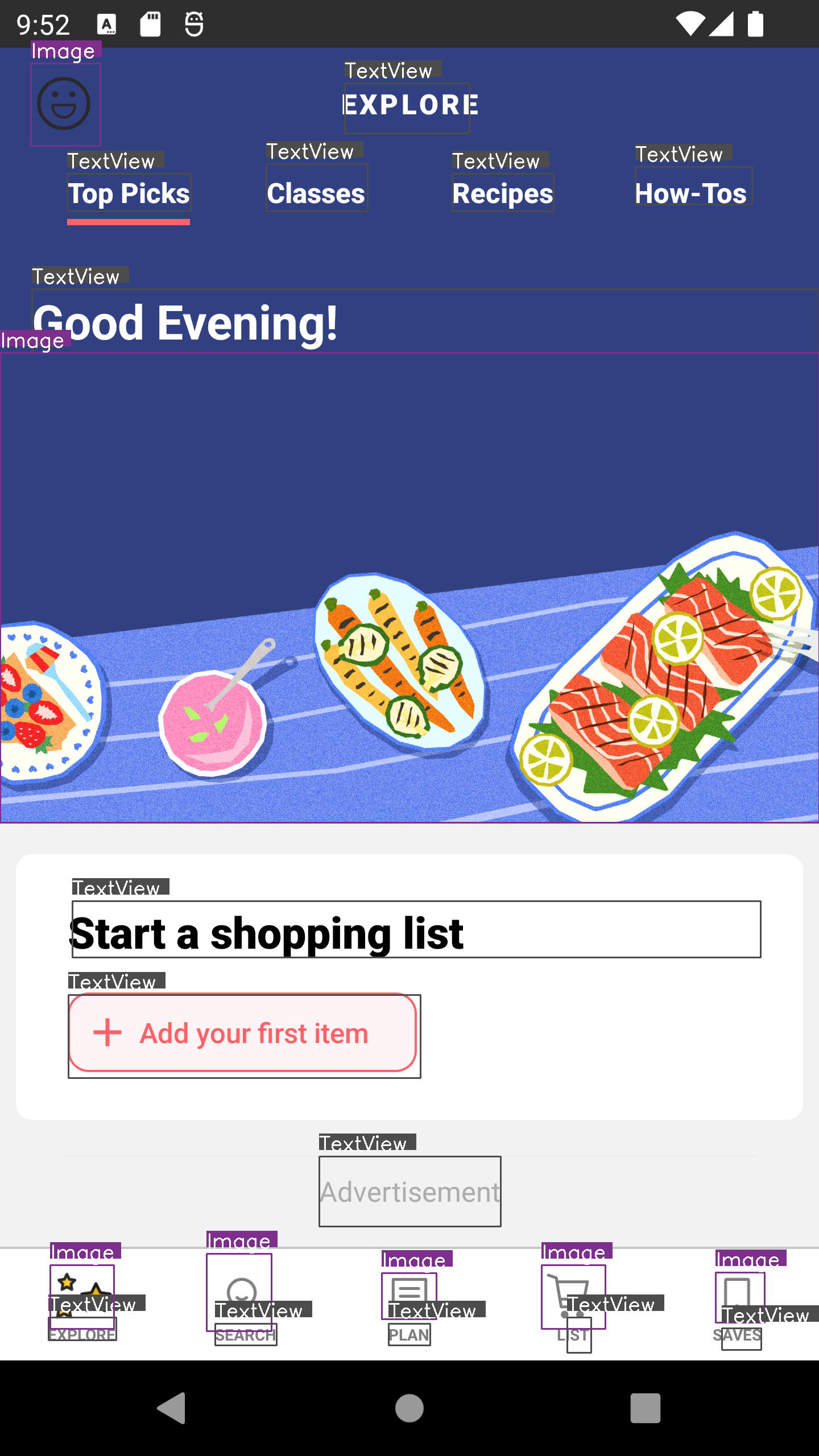}
		\label{fig:vis_16196}}
	\caption{Examples of the element detection in \tool dataset.}
	\label{fig:elemenet_dectection}
\end{figure*}

\subsection{Element Detection}
\label{sec:downstream_1}
UI element detection is a specialized task within object detection, which involves identifying the locations and types of UI elements from a screenshot. It can be useful as a standalone task or as an initial step for more advanced UI modeling.

\textbf{Experimental Setup.}
We conduct experiments using Faster-RCNN\cite{ren2015faster}, one of the highest-performing object detection models evaluated on public datasets.
The model employs a convolutional neural network (CNN) to extract image features from the input UI screenshot. 
It then uses a region proposal network (RPN) to generate region proposals that likely contain an object of interest, as opposed to just background. 
These region proposals are referred to as regions of interest.
Finally, it utilizes an object detection network that predicts object classification scores for the region proposals and determines the object bounding box.

Regarding our training/testing data split, a simple random split would not effectively evaluate the model's generalizability, as screens within the same app may exhibit very similar visual appearances.
To avoid this data leakage issue~\cite{kaufman2012leakage}, we split the screens in the dataset by apps, ensuring that representations of app categories are similar in both the training and testing datasets. The resulting split includes 12,471 (70\%) UIs in the training dataset, 2,771 (15\%) in the validation dataset, and 2,890 (15\%) in the testing dataset.
We adopt the configuration from previous work~\cite{chen2020object} to train the model, employing early stopping on the validation metric to reduce the risk of overfitting. 
We also train a baseline model following the same configuration but use the entire Rico dataset for training and our testing dataset for evaluation.

We assess our model's performance using Average Precision (AP), a standard evaluation metric for object detection. 
We select a threshold of > 0.5 IoU (Intersection over Union), a common benchmark used in object detection challenges~\cite{everingham2010pascal}, to pair a detection with a ground truth UI element.

\textbf{Results.}
Table~\ref{tab:element_detection} presents the Average Precision (AP) at an Intersection over Union (IoU) of 0.5 for each of the 9 classes trained in the Rico and \tool datasets.
We can see that \tool achieves a higher AP across all 9 classes and boosts 10.5\% on the average AP over the performance of Rico.
Fig.~\ref{fig:elemenet_dectection} provides examples of detection results from our \tool.
This superior performance could be attributed to two potential factors.
First, the elements in the Rico dataset may be outdated, thereby training on old-fashioned UIs hinders its capacity to detect more modern elements and poses challenges to modeling contemporary UIs.
Second, the noisy view hierarchy in the Rico dataset may significantly impact the performance of element detection.
In contrast, our \tool dataset features a clean view hierarchy, which enhances the detector's ability to recognize elements of interest, resulting in improved performance.

\begin{figure*}
	\centering
	\includegraphics[width = 0.68\linewidth]{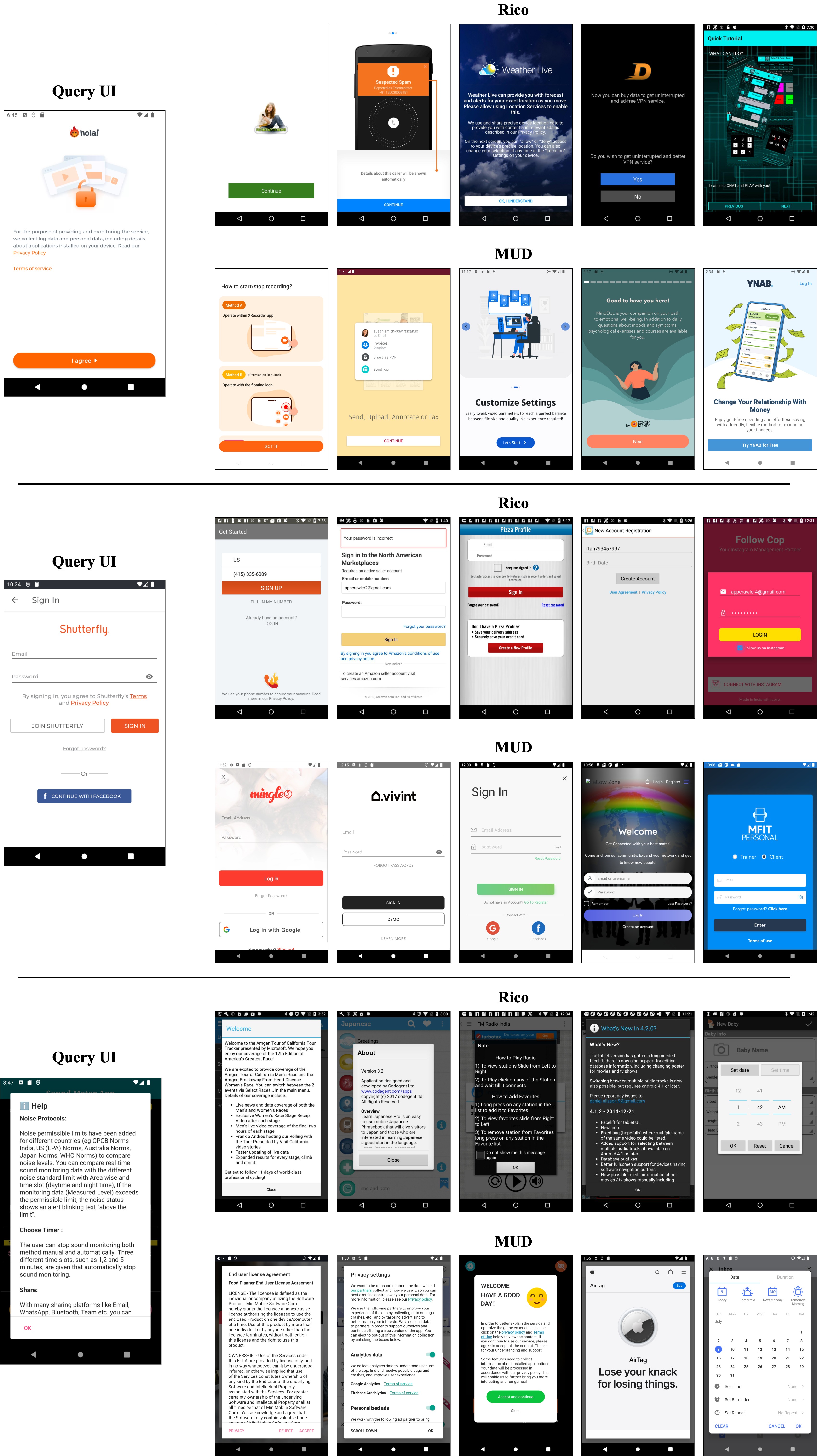}
	\caption{Examples of query results from Rico and \tool datasets.}
	\label{fig:retrival}
\end{figure*}

\subsection{UI Retrieval}
Searching for relevant UI design examples can assist designers in drawing inspiration and comparing design alternatives~\cite{bunian2021vins}.
This task emphasizes the similarity between UI screens, specifically, identifying the top-N most similar screens in the dataset.


\textbf{Experimental Setup.}
We conduct experiments using the autoencoder model proposed in previous work ~\cite{deka2017rico}, which supports unsupervised learning of lower-dimensional representations of the UI.
The model first extracts the layout of a UI using the bounding boxes of all the UI elements in the view hierarchy, distinguishing between text and non-text elements using different colors. 
Then, the model employs a typical encoder-decoder architecture, where the encoder maps the layout image to a lower-dimensional vector, while the decoder is optimized to map this lower-dimensional vector back to the original image.
In detail, the encoder has an input dimension of 11,200, followed by two hidden layers of sizes 2,048 and 256, with an output dimension of size 64. The decoder has the reverse architecture, i.e., from 64 to 256, 2,048, and then 11,200.
We use the mean squared error (MSE) between the input of the encoder and the output of the decoder to train the model. 

We recruit three participants with backgrounds in UI/UX design and art practice to evaluate the performance of UI design retrieval.
At the beginning of the experiment, we provide them with an introduction to our study and ask them to review previous studies~\cite{bunian2021vins,huang2019swire} to better understand the principles of UI retrieval. 
Subsequently, we randomly select 5 query UIs in our dataset and for each query, we use our model to retrieve the top-5 most similar UIs.
Similarly, we retrieve the top-5 most similar UIs from the Rico dataset as the baseline.
The participants then individually rate the results on a five-point Likert scale (1: not related at all, 5: highly related) to assess their relevance to the query.
Note that they do not know which result is from Rico or \tool dataset.

\textbf{Results.}
Fig.~\ref{fig:retrival} shows several example query UIs and their retrieved designs from Rico and \tool datasets.
All participants note that the retrieved results are relevant to the input queries and would serve as valuable design examples. 
As a result, participants assign our dataset an average relevance score of 4.1 out of 5.0, compared to a score of 3.2 for the Rico dataset.
This indicates the potential usefulness of our dataset in retrieving similar UI designs and inspiring creativity.
One participant underscores the key contribution of our dataset, i.e., the inclusion of modern UI designs, stating: \textit{``As a designer, I frequently need to keep up with the latest trends in UI design. However, the UI designs on the existing design-sharing platforms like Dribbble are often more artistic than practical, which doesn't help in gaining 'practical' inspiration. On the other hand, the existing UI design repositories like Rico are outdated and often lack the updated features and functionalities required for contemporary user interface design. I appreciate this new UI design repository. And hope it could keep updating.''}

\section{Discussion}
We analyze the effectiveness of the proposed automated app exploration approach in Section~\ref{sec:data_analysis} and explore the usefulness of the collected UI dataset \tool in Section~\ref{sec:downstream}.
In this section, we delve deeper into the implications of our research, the limitations of our approach, and how future studies can expand upon our work.

\subsection{Improving Automated App Exploration}
The advantages of LLMs in performing semantically-driven app explorations are discussed in Section~\ref{sec:exploration_coverage}, including valid text input, complex compound actions, and multilingual content understanding.
These advantages might due to the knowledge from extensive instructional resources like WikiHow~\cite{web:wikihow} and PixelHelp~\cite{web:pixelhelp}, which provide step-by-step app instructions.
While we use the gpt-3.5-turbo as our model in the study, we believe that other LLMs trained on similar resources, such as the PaLM~\cite{chowdhery2022palm} and the open-sourced LLama model~\cite{touvron2023llama}, could also deliver comparable or even better performance.
In our future work, we aim to conduct a comparative analysis of these LLMs' performance to gain insights into each model's unique strengths and limitations.
To prompt potential exploration guidance, We utilize the LLMs by providing UI view hierarchy information as the context.
However, mobile apps contain multiple sources of information, including app descriptions and functionality introductions. 
Such information could potentially assist our tool in prioritizing the exploration of the most crucial UIs in the apps, thereby enhancing the efficiency of UI collection.
Another limitation of our investigation is our exclusive use of view hierarchy information as the context, while other information remains unused.
For instance, a historical context of previously explored UI states could help LLMs reduce redundant exploration. 
We also hypothesize that providing the context of the app name may improve our approach's performance, as LLMs exposure to tutorials containing step-by-step instructions or descriptions on how to trigger certain features in the training corpus. 
Future research could incorporate these information sources as context for the LLMs, potentially leading to more meaningful and efficient exploration.

Apart from the improvement of LLMs, the automated app exploration may limit to the login activity.
While we propose a novel login approach by Google account authentication that allows 100\% successful login into the Google account to unlock apps, a few apps may only provide customized registration.
The LLMs might simulate account details such as username and password.
While these details appear reasonable, apps often require account activation through third-party verification like email, messaging, or CAPTCHA. 
This results in registration failure, hindering the app exploration.
In the future work, we plan to systematicallty investiagte these failure and develop a complete registration/login system, i.e., monitoring the incoming activation message, activating with proposed steps, sending feedback to login the app. 

Besides enhancing LLMs, the automated app exploration may be restricted to the login activity. 
We propose a novel automatic login approach using Google account authentication, which ensures 100\% successful login into Google accounts to unlock apps. 
However, some apps might only offer customized registration. 
The LLMs could simulate account details like usernames and passwords. 
While these details seem reasonable, apps often require account activation through third-party verification such as email, messaging, or CAPTCHA.
This leads to registration failure, thereby obstructing the app exploration.
In the future, we aim to systematically investigate these failures and design a comprehensive registration/login system.
This would involve monitoring the incoming activation message, activating with proposed steps, and providing feedback to log into the app.

\subsection{Increasing the Dataset}
As data-driven computational methods for modeling UIs continue to rise, datasets have become a pivotal resource for understanding UI. 
In line with this, we have been collecting and annotating the \tool dataset, a large mobile UI dataset comprising modern UI screens and high-quality view hierarchies. 
One limitation of our dataset is that it only captures screenshots and view hierarchies. 
Other artifacts, such as interaction traces and UI animations, could also prove valuable for UI understanding.
Although we captured these artifacts during our automated app exploration, they can appear noisy due to our process of UI removal, rather than their intrinsic characteristics.
For instance, the corresponding UIs in the interaction trace could be significantly impacted by the state duplication.
We plan to investigate this issue and propose an automated algorithm to refactor the trace, thereby broadening UI modeling opportunities for future research.
Aside from multi-modal artifacts, our dataset only collects UIs from Android apps.
Constructing a dataset from different platforms, such as iOS and the web, could yield similar benefits. 
We anticipate that our automated app exploration approach, which relies on view hierarchy for guidance, could be extended to these platforms, given their similar view hierarchy data or equivalence. 
In the future, we aim to broaden our approach to collecting UI datasets across various platforms to enhance the research on UI understanding in a more general context.

\subsection{Improving UI Understanding}
Our experiments have initially showcased the utility of \tool in two common downstream tasks as described in Section~\ref{sec:downstream}.
A logical progression would be to evaluate the performance of our dataset in other UI modeling tasks, such as screen summarization~\cite{wang2021screen2words}, screen question-answering~\cite{wang2023enabling}, etc. 
We expect that these models can attain relative performance enhancements consistent with those findings in Section~\ref{sec:downstream}. 
The implications of these improvements may also shed light on the progression of UIs over the years, potentially serving as a proactive mechanism to remind researchers of advancements in UI understanding.
Beyond that, a long-term objective of our data collection and high-quality human annotation efforts is to achieve a more generalized and modern understanding of UIs. 
This would involve developing an advanced model capable of comprehending the latest UI designs and their semantics, ultimately enabling an intelligent UI agent to tackle various UI problems.
\section{Conclusion}
In this paper, we present \tool, a dataset of 18k UIs paired with visual and hierarchy information to aid UI modeling. 
Unlike most existing datasets for UI research, which often lack sufficient data or contain noisy and erroneous data, \tool is collected using a novel approach that incorporates best practices. 
In detail, we employ the Large Language Models (LLMs) to simulate an app expert for automatically exploring apps, thereby collecting an order of magnitude more UIs. 
Subsequently, we utilize the best practices of UI noise filtering techniques to eliminate noise and further implement human annotation to ensure the final quality of the dataset. 
We demonstrate the effectiveness of LLMs-enhanced automated app exploration in generating human-like actions, facilitating more thorough and efficient app exploration compared to three state-of-the-art tools. 
Moreover, we highlight the utility of our dataset \tool by modeling two common UI tasks: element detection and UI retrieval. 
The dataset \tool will be released to the public to encourage further research and lay the groundwork for UI modeling based on high-quality and modern UI designs.

\begin{acks}
	We appreciate the assistance provided by Will Middlewick in facilitating the UI collection process, as well as the students from Monash University who have contributed to conducting empirical studies and some experiments.
\end{acks}

\bibliographystyle{ACM-Reference-Format}
\bibliography{main}


\begin{thebibliography}{76}


\ifx \showCODEN    \undefined \def \showCODEN     #1{\unskip}     \fi
\ifx \showDOI      \undefined \def \showDOI       #1{#1}\fi
\ifx \showISBNx    \undefined \def \showISBNx     #1{\unskip}     \fi
\ifx \showISBNxiii \undefined \def \showISBNxiii  #1{\unskip}     \fi
\ifx \showISSN     \undefined \def \showISSN      #1{\unskip}     \fi
\ifx \showLCCN     \undefined \def \showLCCN      #1{\unskip}     \fi
\ifx \shownote     \undefined \def \shownote      #1{#1}          \fi
\ifx \showarticletitle \undefined \def \showarticletitle #1{#1}   \fi
\ifx \showURL      \undefined \def \showURL       {\relax}        \fi
\providecommand\bibfield[2]{#2}
\providecommand\bibinfo[2]{#2}
\providecommand\natexlab[1]{#1}
\providecommand\showeprint[2][]{arXiv:#2}

\bibitem[web(2023a)]%
        {web:adb}
 \bibinfo{year}{2023}\natexlab{a}.
\newblock \bibinfo{title}{Android Debug Bridge (adb) - Android Developers}.
\newblock
  \bibinfo{howpublished}{\url{https://developer.android.com/studio/command-line/adb}}.
\newblock


\bibitem[web(2023b)]%
        {web:uiautomator}
 \bibinfo{year}{2023}\natexlab{b}.
\newblock \bibinfo{title}{Android Uiautomator2 Python Wrapper}.
\newblock
  \bibinfo{howpublished}{\url{https://github.com/openatx/uiautomator2}}.
\newblock


\bibitem[web(2023c)]%
        {web:chromedriver}
 \bibinfo{year}{2023}\natexlab{c}.
\newblock \bibinfo{title}{ChromeDriver - WebDriver for Chrome}.
\newblock \bibinfo{howpublished}{\url{https://chromedriver.chromium.org/}}.
\newblock


\bibitem[web(2023d)]%
        {web:language}
 \bibinfo{year}{2023}\natexlab{d}.
\newblock \bibinfo{title}{How Many Languages Does ChatGPT Support? The Complete
  ChatGPT Language List}.
\newblock
  \bibinfo{howpublished}{\url{https://seo.ai/blog/how-many-languages-does-chatgpt-support}}.
\newblock


\bibitem[web(2023e)]%
        {web:chatgpt}
 \bibinfo{year}{2023}\natexlab{e}.
\newblock \bibinfo{title}{Introducing ChatGPT}.
\newblock \bibinfo{howpublished}{\url{https://chat.openai.com/}}.
\newblock


\bibitem[web(2023f)]%
        {web:material}
 \bibinfo{year}{2023}\natexlab{f}.
\newblock \bibinfo{title}{Material Design}.
\newblock \bibinfo{howpublished}{\url{https://m3.material.io/}}.
\newblock


\bibitem[web(2023g)]%
        {web:login}
 \bibinfo{year}{2023}\natexlab{g}.
\newblock \bibinfo{title}{Overview of One Tap sign-in on Android}.
\newblock
  \bibinfo{howpublished}{\url{https://developers.google.com/identity/one-tap/android/overview}}.
\newblock


\bibitem[web(2023h)]%
        {web:pixelhelp}
 \bibinfo{year}{2023}\natexlab{h}.
\newblock \bibinfo{title}{Pixel Phone Help}.
\newblock \bibinfo{howpublished}{\url{https://support.google.com/pixelphone/}}.
\newblock


\bibitem[web(2023i)]%
        {web:toast}
 \bibinfo{year}{2023}\natexlab{i}.
\newblock \bibinfo{title}{Toasts overview}.
\newblock
  \bibinfo{howpublished}{\url{https://developer.android.com/guide/topics/ui/notifiers/toasts}}.
\newblock


\bibitem[web(2023j)]%
        {web:monkey}
 \bibinfo{year}{2023}\natexlab{j}.
\newblock \bibinfo{title}{UI/Application Exerciser Monkey}.
\newblock
  \bibinfo{howpublished}{\url{https://developer.android.com/studio/test/other-testing-tools/monkey}}.
\newblock


\bibitem[web(2023k)]%
        {web:wikihow}
 \bibinfo{year}{2023}\natexlab{k}.
\newblock \bibinfo{title}{wikiHow: How-to instructions you can trust.}
\newblock \bibinfo{howpublished}{\url{https://www.wikihow.com/Main-Page}}.
\newblock


\bibitem[Amalfitano et~al\mbox{.}(2012)]%
        {amalfitano2012using}
\bibfield{author}{\bibinfo{person}{Domenico Amalfitano},
  \bibinfo{person}{Anna~Rita Fasolino}, \bibinfo{person}{Porfirio Tramontana},
  \bibinfo{person}{Salvatore De~Carmine}, {and} \bibinfo{person}{Atif~M
  Memon}.} \bibinfo{year}{2012}\natexlab{}.
\newblock \showarticletitle{Using GUI ripping for automated testing of Android
  applications}. In \bibinfo{booktitle}{\emph{2012 Proceedings of the 27th
  IEEE/ACM International Conference on Automated Software Engineering}}. IEEE,
  \bibinfo{pages}{258--261}.
\newblock


\bibitem[Bernal-C{\'a}rdenas et~al\mbox{.}(2019)]%
        {bernal2019guigle}
\bibfield{author}{\bibinfo{person}{Carlos Bernal-C{\'a}rdenas},
  \bibinfo{person}{Kevin Moran}, \bibinfo{person}{Michele Tufano},
  \bibinfo{person}{Zichang Liu}, \bibinfo{person}{Linyong Nan},
  \bibinfo{person}{Zhehan Shi}, {and} \bibinfo{person}{Denys Poshyvanyk}.}
  \bibinfo{year}{2019}\natexlab{}.
\newblock \showarticletitle{Guigle: A gui search engine for android apps}. In
  \bibinfo{booktitle}{\emph{2019 IEEE/ACM 41st International Conference on
  Software Engineering: Companion Proceedings (ICSE-Companion)}}. IEEE,
  \bibinfo{pages}{71--74}.
\newblock


\bibitem[Brown et~al\mbox{.}(2020)]%
        {brown2020language}
\bibfield{author}{\bibinfo{person}{Tom Brown}, \bibinfo{person}{Benjamin Mann},
  \bibinfo{person}{Nick Ryder}, \bibinfo{person}{Melanie Subbiah},
  \bibinfo{person}{Jared~D Kaplan}, \bibinfo{person}{Prafulla Dhariwal},
  \bibinfo{person}{Arvind Neelakantan}, \bibinfo{person}{Pranav Shyam},
  \bibinfo{person}{Girish Sastry}, \bibinfo{person}{Amanda Askell},
  {et~al\mbox{.}}} \bibinfo{year}{2020}\natexlab{}.
\newblock \showarticletitle{Language models are few-shot learners}.
\newblock \bibinfo{journal}{\emph{Advances in neural information processing
  systems}}  \bibinfo{volume}{33} (\bibinfo{year}{2020}),
  \bibinfo{pages}{1877--1901}.
\newblock


\bibitem[Bunian et~al\mbox{.}(2021)]%
        {bunian2021vins}
\bibfield{author}{\bibinfo{person}{Sara Bunian}, \bibinfo{person}{Kai Li},
  \bibinfo{person}{Chaima Jemmali}, \bibinfo{person}{Casper Harteveld},
  \bibinfo{person}{Yun Fu}, {and} \bibinfo{person}{Magy~Seif Seif El-Nasr}.}
  \bibinfo{year}{2021}\natexlab{}.
\newblock \showarticletitle{Vins: Visual search for mobile user interface
  design}. In \bibinfo{booktitle}{\emph{Proceedings of the 2021 CHI Conference
  on Human Factors in Computing Systems}}. \bibinfo{pages}{1--14}.
\newblock


\bibitem[Chen et~al\mbox{.}(2020a)]%
        {chen2020lost}
\bibfield{author}{\bibinfo{person}{Chunyang Chen}, \bibinfo{person}{Sidong
  Feng}, \bibinfo{person}{Zhengyang Liu}, \bibinfo{person}{Zhenchang Xing},
  {and} \bibinfo{person}{Shengdong Zhao}.} \bibinfo{year}{2020}\natexlab{a}.
\newblock \showarticletitle{From lost to found: Discover missing ui design
  semantics through recovering missing tags}.
\newblock \bibinfo{journal}{\emph{Proceedings of the ACM on Human-Computer
  Interaction}} \bibinfo{volume}{4}, \bibinfo{number}{CSCW2}
  (\bibinfo{year}{2020}), \bibinfo{pages}{1--22}.
\newblock


\bibitem[Chen et~al\mbox{.}(2019b)]%
        {chen2019gallery}
\bibfield{author}{\bibinfo{person}{Chunyang Chen}, \bibinfo{person}{Sidong
  Feng}, \bibinfo{person}{Zhenchang Xing}, \bibinfo{person}{Linda Liu},
  \bibinfo{person}{Shengdong Zhao}, {and} \bibinfo{person}{Jinshui Wang}.}
  \bibinfo{year}{2019}\natexlab{b}.
\newblock \showarticletitle{Gallery dc: Design search and knowledge discovery
  through auto-created gui component gallery}.
\newblock \bibinfo{journal}{\emph{Proceedings of the ACM on Human-Computer
  Interaction}} \bibinfo{volume}{3}, \bibinfo{number}{CSCW}
  (\bibinfo{year}{2019}), \bibinfo{pages}{1--22}.
\newblock


\bibitem[Chen et~al\mbox{.}(2018)]%
        {chen2018ui}
\bibfield{author}{\bibinfo{person}{Chunyang Chen}, \bibinfo{person}{Ting Su},
  \bibinfo{person}{Guozhu Meng}, \bibinfo{person}{Zhenchang Xing}, {and}
  \bibinfo{person}{Yang Liu}.} \bibinfo{year}{2018}\natexlab{}.
\newblock \showarticletitle{From ui design image to gui skeleton: a neural
  machine translator to bootstrap mobile gui implementation}. In
  \bibinfo{booktitle}{\emph{Proceedings of the 40th International Conference on
  Software Engineering}}. \bibinfo{pages}{665--676}.
\newblock


\bibitem[Chen et~al\mbox{.}(2023)]%
        {chen2023unveiling}
\bibfield{author}{\bibinfo{person}{Jieshan Chen}, \bibinfo{person}{Jiamou Sun},
  \bibinfo{person}{Sidong Feng}, \bibinfo{person}{Zhenchang Xing},
  \bibinfo{person}{Qinghua Lu}, \bibinfo{person}{Xiwei Xu}, {and}
  \bibinfo{person}{Chunyang Chen}.} \bibinfo{year}{2023}\natexlab{}.
\newblock \showarticletitle{Unveiling the Tricks: Automated Detection of Dark
  Patterns in Mobile Applications}. In \bibinfo{booktitle}{\emph{Proceedings of
  the 36th Annual ACM Symposium on User Interface Software and Technology}}.
  \bibinfo{pages}{1--20}.
\newblock


\bibitem[Chen et~al\mbox{.}(2020b)]%
        {chen2020object}
\bibfield{author}{\bibinfo{person}{Jieshan Chen}, \bibinfo{person}{Mulong Xie},
  \bibinfo{person}{Zhenchang Xing}, \bibinfo{person}{Chunyang Chen},
  \bibinfo{person}{Xiwei Xu}, \bibinfo{person}{Liming Zhu}, {and}
  \bibinfo{person}{Guoqiang Li}.} \bibinfo{year}{2020}\natexlab{b}.
\newblock \showarticletitle{Object detection for graphical user interface: Old
  fashioned or deep learning or a combination?}. In
  \bibinfo{booktitle}{\emph{proceedings of the 28th ACM joint meeting on
  European Software Engineering Conference and Symposium on the Foundations of
  Software Engineering}}. \bibinfo{pages}{1202--1214}.
\newblock


\bibitem[Chen et~al\mbox{.}(2019a)]%
        {chen2019storydroid}
\bibfield{author}{\bibinfo{person}{Sen Chen}, \bibinfo{person}{Lingling Fan},
  \bibinfo{person}{Chunyang Chen}, \bibinfo{person}{Ting Su},
  \bibinfo{person}{Wenhe Li}, \bibinfo{person}{Yang Liu}, {and}
  \bibinfo{person}{Lihua Xu}.} \bibinfo{year}{2019}\natexlab{a}.
\newblock \showarticletitle{Storydroid: Automated generation of storyboard for
  Android apps}. In \bibinfo{booktitle}{\emph{2019 IEEE/ACM 41st International
  Conference on Software Engineering (ICSE)}}. IEEE, \bibinfo{pages}{596--607}.
\newblock


\bibitem[Chowdhery et~al\mbox{.}(2022)]%
        {chowdhery2022palm}
\bibfield{author}{\bibinfo{person}{Aakanksha Chowdhery},
  \bibinfo{person}{Sharan Narang}, \bibinfo{person}{Jacob Devlin},
  \bibinfo{person}{Maarten Bosma}, \bibinfo{person}{Gaurav Mishra},
  \bibinfo{person}{Adam Roberts}, \bibinfo{person}{Paul Barham},
  \bibinfo{person}{Hyung~Won Chung}, \bibinfo{person}{Charles Sutton},
  \bibinfo{person}{Sebastian Gehrmann}, {et~al\mbox{.}}}
  \bibinfo{year}{2022}\natexlab{}.
\newblock \showarticletitle{Palm: Scaling language modeling with pathways}.
\newblock \bibinfo{journal}{\emph{arXiv preprint arXiv:2204.02311}}
  (\bibinfo{year}{2022}).
\newblock


\bibitem[Degott et~al\mbox{.}(2019)]%
        {degott2019learning}
\bibfield{author}{\bibinfo{person}{Christian Degott},
  \bibinfo{person}{Nataniel~P Borges~Jr}, {and} \bibinfo{person}{Andreas
  Zeller}.} \bibinfo{year}{2019}\natexlab{}.
\newblock \showarticletitle{Learning user interface element interactions}. In
  \bibinfo{booktitle}{\emph{Proceedings of the 28th ACM SIGSOFT International
  Symposium on Software Testing and Analysis}}. \bibinfo{pages}{296--306}.
\newblock


\bibitem[Deka et~al\mbox{.}(2017)]%
        {deka2017rico}
\bibfield{author}{\bibinfo{person}{Biplab Deka}, \bibinfo{person}{Zifeng
  Huang}, \bibinfo{person}{Chad Franzen}, \bibinfo{person}{Joshua Hibschman},
  \bibinfo{person}{Daniel Afergan}, \bibinfo{person}{Yang Li},
  \bibinfo{person}{Jeffrey Nichols}, {and} \bibinfo{person}{Ranjitha Kumar}.}
  \bibinfo{year}{2017}\natexlab{}.
\newblock \showarticletitle{Rico: A mobile app dataset for building data-driven
  design applications}. In \bibinfo{booktitle}{\emph{Proceedings of the 30th
  Annual ACM Symposium on User Interface Software and Technology}}.
  \bibinfo{pages}{845--854}.
\newblock


\bibitem[Deka et~al\mbox{.}(2016)]%
        {deka2016erica}
\bibfield{author}{\bibinfo{person}{Biplab Deka}, \bibinfo{person}{Zifeng
  Huang}, {and} \bibinfo{person}{Ranjitha Kumar}.}
  \bibinfo{year}{2016}\natexlab{}.
\newblock \showarticletitle{ERICA: Interaction mining mobile apps}. In
  \bibinfo{booktitle}{\emph{Proceedings of the 29th annual symposium on user
  interface software and technology}}. \bibinfo{pages}{767--776}.
\newblock


\bibitem[Dixon and Fogarty(2010)]%
        {dixon2010prefab}
\bibfield{author}{\bibinfo{person}{Morgan Dixon} {and} \bibinfo{person}{James
  Fogarty}.} \bibinfo{year}{2010}\natexlab{}.
\newblock \showarticletitle{Prefab: implementing advanced behaviors using
  pixel-based reverse engineering of interface structure}. In
  \bibinfo{booktitle}{\emph{Proceedings of the SIGCHI Conference on Human
  Factors in Computing Systems}}. \bibinfo{pages}{1525--1534}.
\newblock


\bibitem[Everingham et~al\mbox{.}(2010)]%
        {everingham2010pascal}
\bibfield{author}{\bibinfo{person}{Mark Everingham}, \bibinfo{person}{Luc
  Van~Gool}, \bibinfo{person}{Christopher~KI Williams}, \bibinfo{person}{John
  Winn}, {and} \bibinfo{person}{Andrew Zisserman}.}
  \bibinfo{year}{2010}\natexlab{}.
\newblock \showarticletitle{The pascal visual object classes (voc) challenge}.
\newblock \bibinfo{journal}{\emph{International journal of computer vision}}
  \bibinfo{volume}{88} (\bibinfo{year}{2010}), \bibinfo{pages}{303--338}.
\newblock


\bibitem[Feng and Chen(2022a)]%
        {feng2022gifdroid1}
\bibfield{author}{\bibinfo{person}{Sidong Feng} {and} \bibinfo{person}{Chunyang
  Chen}.} \bibinfo{year}{2022}\natexlab{a}.
\newblock \showarticletitle{GIFdroid: an automated light-weight tool for
  replaying visual bug reports}. In \bibinfo{booktitle}{\emph{Proceedings of
  the ACM/IEEE 44th International Conference on Software Engineering: Companion
  Proceedings}}. \bibinfo{pages}{95--99}.
\newblock


\bibitem[Feng and Chen(2022b)]%
        {feng2022gifdroid}
\bibfield{author}{\bibinfo{person}{Sidong Feng} {and} \bibinfo{person}{Chunyang
  Chen}.} \bibinfo{year}{2022}\natexlab{b}.
\newblock \showarticletitle{GIFdroid: automated replay of visual bug reports
  for Android apps}. In \bibinfo{booktitle}{\emph{Proceedings of the 44th
  International Conference on Software Engineering}}.
  \bibinfo{pages}{1045--1057}.
\newblock


\bibitem[Feng and Chen(2023)]%
        {feng2023prompting}
\bibfield{author}{\bibinfo{person}{Sidong Feng} {and} \bibinfo{person}{Chunyang
  Chen}.} \bibinfo{year}{2023}\natexlab{}.
\newblock \showarticletitle{Prompting Is All You Need: Automated Android Bug
  Replay with Large Language Models}.
\newblock \bibinfo{journal}{\emph{arXiv preprint arXiv:2306.01987}}
  (\bibinfo{year}{2023}).
\newblock


\bibitem[Feng et~al\mbox{.}(2022a)]%
        {feng2022gallery}
\bibfield{author}{\bibinfo{person}{Sidong Feng}, \bibinfo{person}{Chunyang
  Chen}, {and} \bibinfo{person}{Zhenchang Xing}.}
  \bibinfo{year}{2022}\natexlab{a}.
\newblock \showarticletitle{Gallery DC: Auto-created GUI component gallery for
  design search and knowledge discovery}. In
  \bibinfo{booktitle}{\emph{Proceedings of the ACM/IEEE 44th International
  Conference on Software Engineering: Companion Proceedings}}.
  \bibinfo{pages}{80--84}.
\newblock


\bibitem[Feng et~al\mbox{.}(2023a)]%
        {feng2023video2action}
\bibfield{author}{\bibinfo{person}{Sidong Feng}, \bibinfo{person}{Chunyang
  Chen}, {and} \bibinfo{person}{Zhenchang Xing}.}
  \bibinfo{year}{2023}\natexlab{a}.
\newblock \showarticletitle{Video2Action: Reducing human interactions in action
  annotation of app tutorial videos}. In \bibinfo{booktitle}{\emph{Proceedings
  of the 36th Annual ACM Symposium on User Interface Software and Technology}}.
  \bibinfo{pages}{1--15}.
\newblock


\bibitem[Feng et~al\mbox{.}(2022b)]%
        {feng2022auto}
\bibfield{author}{\bibinfo{person}{Sidong Feng}, \bibinfo{person}{Minmin
  Jiang}, \bibinfo{person}{Tingting Zhou}, \bibinfo{person}{Yankun Zhen}, {and}
  \bibinfo{person}{Chunyang Chen}.} \bibinfo{year}{2022}\natexlab{b}.
\newblock \showarticletitle{Auto-Icon+: An Automated End-to-End Code Generation
  Tool for Icon Designs in UI Development}.
\newblock \bibinfo{journal}{\emph{ACM Transactions on Interactive Intelligent
  Systems}} \bibinfo{volume}{12}, \bibinfo{number}{4} (\bibinfo{year}{2022}),
  \bibinfo{pages}{1--26}.
\newblock


\bibitem[Feng et~al\mbox{.}(2023b)]%
        {feng2023towards}
\bibfield{author}{\bibinfo{person}{Sidong Feng}, \bibinfo{person}{Haochuan Lu},
  \bibinfo{person}{Ting Xiong}, \bibinfo{person}{Yuetang Deng}, {and}
  \bibinfo{person}{Chunyang Chen}.} \bibinfo{year}{2023}\natexlab{b}.
\newblock \showarticletitle{Towards Efficient Record and Replay: A Case Study
  in WeChat}. In \bibinfo{booktitle}{\emph{Proceedings of the 31st ACM Joint
  European Software Engineering Conference and Symposium on the Foundations of
  Software Engineering}}. \bibinfo{pages}{1681--1692}.
\newblock


\bibitem[Feng et~al\mbox{.}(2021)]%
        {feng2021auto}
\bibfield{author}{\bibinfo{person}{Sidong Feng}, \bibinfo{person}{Suyu Ma},
  \bibinfo{person}{Jinzhong Yu}, \bibinfo{person}{Chunyang Chen},
  \bibinfo{person}{Tingting Zhou}, {and} \bibinfo{person}{Yankun Zhen}.}
  \bibinfo{year}{2021}\natexlab{}.
\newblock \showarticletitle{Auto-icon: An automated code generation tool for
  icon designs assisting in ui development}. In \bibinfo{booktitle}{\emph{26th
  International Conference on Intelligent User Interfaces}}.
  \bibinfo{pages}{59--69}.
\newblock


\bibitem[Feng et~al\mbox{.}(2023c)]%
        {feng2023efficiency}
\bibfield{author}{\bibinfo{person}{Sidong Feng}, \bibinfo{person}{Mulong Xie},
  {and} \bibinfo{person}{Chunyang Chen}.} \bibinfo{year}{2023}\natexlab{c}.
\newblock \showarticletitle{Efficiency matters: Speeding up automated testing
  with gui rendering inference}. In \bibinfo{booktitle}{\emph{2023 IEEE/ACM
  45th International Conference on Software Engineering (ICSE)}}. IEEE,
  \bibinfo{pages}{906--918}.
\newblock


\bibitem[Feng et~al\mbox{.}(2023d)]%
        {feng2023read}
\bibfield{author}{\bibinfo{person}{Sidong Feng}, \bibinfo{person}{Mulong Xie},
  \bibinfo{person}{Yinxing Xue}, {and} \bibinfo{person}{Chunyang Chen}.}
  \bibinfo{year}{2023}\natexlab{d}.
\newblock \showarticletitle{Read It, Don't Watch It: Captioning Bug Recordings
  Automatically}.
\newblock \bibinfo{journal}{\emph{arXiv preprint arXiv:2302.00886}}
  (\bibinfo{year}{2023}).
\newblock


\bibitem[Feng et~al\mbox{.}(2023e)]%
        {feng2023designing}
\bibfield{author}{\bibinfo{person}{Sidong Feng}, \bibinfo{person}{Mingyue
  Yuan}, \bibinfo{person}{Jieshan Chen}, \bibinfo{person}{Zhenchang Xing},
  {and} \bibinfo{person}{Chunyang Chen}.} \bibinfo{year}{2023}\natexlab{e}.
\newblock \showarticletitle{Designing with Language: Wireframing UI Design
  Intent with Generative Large Language Models}.
\newblock \bibinfo{journal}{\emph{arXiv preprint arXiv:2312.07755}}
  (\bibinfo{year}{2023}).
\newblock


\bibitem[Fok et~al\mbox{.}(2022)]%
        {fok2022large}
\bibfield{author}{\bibinfo{person}{Raymond Fok}, \bibinfo{person}{Mingyuan
  Zhong}, \bibinfo{person}{Anne~Spencer Ross}, \bibinfo{person}{James Fogarty},
  {and} \bibinfo{person}{Jacob~O Wobbrock}.} \bibinfo{year}{2022}\natexlab{}.
\newblock \showarticletitle{A Large-Scale Longitudinal Analysis of Missing
  Label Accessibility Failures in Android Apps}. In
  \bibinfo{booktitle}{\emph{Proceedings of the 2022 CHI Conference on Human
  Factors in Computing Systems}}. \bibinfo{pages}{1--16}.
\newblock


\bibitem[Fraher and Boyd-Brent(2010)]%
        {fraher2010gestalt}
\bibfield{author}{\bibinfo{person}{Robert Fraher} {and} \bibinfo{person}{James
  Boyd-Brent}.} \bibinfo{year}{2010}\natexlab{}.
\newblock \showarticletitle{Gestalt theory, engagement and interaction}.
\newblock In \bibinfo{booktitle}{\emph{CHI'10 Extended Abstracts on Human
  Factors in Computing Systems}}. \bibinfo{pages}{3211--3216}.
\newblock


\bibitem[Gu et~al\mbox{.}(2019)]%
        {gu2019practical}
\bibfield{author}{\bibinfo{person}{Tianxiao Gu}, \bibinfo{person}{Chengnian
  Sun}, \bibinfo{person}{Xiaoxing Ma}, \bibinfo{person}{Chun Cao},
  \bibinfo{person}{Chang Xu}, \bibinfo{person}{Yuan Yao},
  \bibinfo{person}{Qirun Zhang}, \bibinfo{person}{Jian Lu}, {and}
  \bibinfo{person}{Zhendong Su}.} \bibinfo{year}{2019}\natexlab{}.
\newblock \showarticletitle{Practical GUI testing of Android applications via
  model abstraction and refinement}. In \bibinfo{booktitle}{\emph{2019 IEEE/ACM
  41st International Conference on Software Engineering (ICSE)}}. IEEE,
  \bibinfo{pages}{269--280}.
\newblock


\bibitem[Huang et~al\mbox{.}(2019)]%
        {huang2019swire}
\bibfield{author}{\bibinfo{person}{Forrest Huang}, \bibinfo{person}{John~F
  Canny}, {and} \bibinfo{person}{Jeffrey Nichols}.}
  \bibinfo{year}{2019}\natexlab{}.
\newblock \showarticletitle{Swire: Sketch-based user interface retrieval}. In
  \bibinfo{booktitle}{\emph{Proceedings of the 2019 CHI Conference on Human
  Factors in Computing Systems}}. \bibinfo{pages}{1--10}.
\newblock


\bibitem[Kaufman et~al\mbox{.}(2012)]%
        {kaufman2012leakage}
\bibfield{author}{\bibinfo{person}{Shachar Kaufman}, \bibinfo{person}{Saharon
  Rosset}, \bibinfo{person}{Claudia Perlich}, {and} \bibinfo{person}{Ori
  Stitelman}.} \bibinfo{year}{2012}\natexlab{}.
\newblock \showarticletitle{Leakage in data mining: Formulation, detection, and
  avoidance}.
\newblock \bibinfo{journal}{\emph{ACM Transactions on Knowledge Discovery from
  Data (TKDD)}} \bibinfo{volume}{6}, \bibinfo{number}{4}
  (\bibinfo{year}{2012}), \bibinfo{pages}{1--21}.
\newblock


\bibitem[Kumar et~al\mbox{.}(2013)]%
        {kumar2013webzeitgeist}
\bibfield{author}{\bibinfo{person}{Ranjitha Kumar}, \bibinfo{person}{Arvind
  Satyanarayan}, \bibinfo{person}{Cesar Torres}, \bibinfo{person}{Maxine Lim},
  \bibinfo{person}{Salman Ahmad}, \bibinfo{person}{Scott~R Klemmer}, {and}
  \bibinfo{person}{Jerry~O Talton}.} \bibinfo{year}{2013}\natexlab{}.
\newblock \showarticletitle{Webzeitgeist: design mining the web}. In
  \bibinfo{booktitle}{\emph{Proceedings of the SIGCHI Conference on Human
  Factors in Computing Systems}}. \bibinfo{pages}{3083--3092}.
\newblock


\bibitem[Leiva et~al\mbox{.}(2020)]%
        {leiva2020enrico}
\bibfield{author}{\bibinfo{person}{Luis~A Leiva}, \bibinfo{person}{Asutosh
  Hota}, {and} \bibinfo{person}{Antti Oulasvirta}.}
  \bibinfo{year}{2020}\natexlab{}.
\newblock \showarticletitle{Enrico: A dataset for topic modeling of mobile UI
  designs}. In \bibinfo{booktitle}{\emph{22nd International Conference on
  Human-Computer Interaction with Mobile Devices and Services}}.
  \bibinfo{pages}{1--4}.
\newblock


\bibitem[Li et~al\mbox{.}(2022)]%
        {li2022learning}
\bibfield{author}{\bibinfo{person}{Gang Li}, \bibinfo{person}{Gilles Baechler},
  \bibinfo{person}{Manuel Tragut}, {and} \bibinfo{person}{Yang Li}.}
  \bibinfo{year}{2022}\natexlab{}.
\newblock \showarticletitle{Learning to denoise raw mobile UI layouts for
  improving datasets at scale}. In \bibinfo{booktitle}{\emph{Proceedings of the
  2022 CHI Conference on Human Factors in Computing Systems}}.
  \bibinfo{pages}{1--13}.
\newblock


\bibitem[Li et~al\mbox{.}(2021)]%
        {li2021screen2vec}
\bibfield{author}{\bibinfo{person}{Toby Jia-Jun Li}, \bibinfo{person}{Lindsay
  Popowski}, \bibinfo{person}{Tom Mitchell}, {and} \bibinfo{person}{Brad~A
  Myers}.} \bibinfo{year}{2021}\natexlab{}.
\newblock \showarticletitle{Screen2vec: Semantic embedding of gui screens and
  gui components}. In \bibinfo{booktitle}{\emph{Proceedings of the 2021 CHI
  Conference on Human Factors in Computing Systems}}. \bibinfo{pages}{1--15}.
\newblock


\bibitem[Li et~al\mbox{.}(2020a)]%
        {li2020mapping}
\bibfield{author}{\bibinfo{person}{Yang Li}, \bibinfo{person}{Jiacong He},
  \bibinfo{person}{Xin Zhou}, \bibinfo{person}{Yuan Zhang}, {and}
  \bibinfo{person}{Jason Baldridge}.} \bibinfo{year}{2020}\natexlab{a}.
\newblock \showarticletitle{Mapping Natural Language Instructions to Mobile UI
  Action Sequences}. In \bibinfo{booktitle}{\emph{Proceedings of the 58th
  Annual Meeting of the Association for Computational Linguistics}}.
  \bibinfo{pages}{8198--8210}.
\newblock


\bibitem[Li et~al\mbox{.}(2020b)]%
        {li2020widget}
\bibfield{author}{\bibinfo{person}{Yang Li}, \bibinfo{person}{Gang Li},
  \bibinfo{person}{Luheng He}, \bibinfo{person}{Jingjie Zheng},
  \bibinfo{person}{Hong Li}, {and} \bibinfo{person}{Zhiwei Guan}.}
  \bibinfo{year}{2020}\natexlab{b}.
\newblock \showarticletitle{Widget captioning: Generating natural language
  description for mobile user interface elements}.
\newblock \bibinfo{journal}{\emph{arXiv preprint arXiv:2010.04295}}
  (\bibinfo{year}{2020}).
\newblock


\bibitem[Li et~al\mbox{.}(2017)]%
        {li2017droidbot}
\bibfield{author}{\bibinfo{person}{Yuanchun Li}, \bibinfo{person}{Ziyue Yang},
  \bibinfo{person}{Yao Guo}, {and} \bibinfo{person}{Xiangqun Chen}.}
  \bibinfo{year}{2017}\natexlab{}.
\newblock \showarticletitle{Droidbot: a lightweight ui-guided test input
  generator for android}. In \bibinfo{booktitle}{\emph{2017 IEEE/ACM 39th
  International Conference on Software Engineering Companion (ICSE-C)}}. IEEE,
  \bibinfo{pages}{23--26}.
\newblock


\bibitem[Li et~al\mbox{.}(2019)]%
        {li2019humanoid}
\bibfield{author}{\bibinfo{person}{Yuanchun Li}, \bibinfo{person}{Ziyue Yang},
  \bibinfo{person}{Yao Guo}, {and} \bibinfo{person}{Xiangqun Chen}.}
  \bibinfo{year}{2019}\natexlab{}.
\newblock \showarticletitle{Humanoid: A deep learning-based approach to
  automated black-box android app testing}. In \bibinfo{booktitle}{\emph{2019
  34th IEEE/ACM International Conference on Automated Software Engineering
  (ASE)}}. IEEE, \bibinfo{pages}{1070--1073}.
\newblock


\bibitem[Liu et~al\mbox{.}(2019)]%
        {liu2019roberta}
\bibfield{author}{\bibinfo{person}{Yinhan Liu}, \bibinfo{person}{Myle Ott},
  \bibinfo{person}{Naman Goyal}, \bibinfo{person}{Jingfei Du},
  \bibinfo{person}{Mandar Joshi}, \bibinfo{person}{Danqi Chen},
  \bibinfo{person}{Omer Levy}, \bibinfo{person}{Mike Lewis},
  \bibinfo{person}{Luke Zettlemoyer}, {and} \bibinfo{person}{Veselin
  Stoyanov}.} \bibinfo{year}{2019}\natexlab{}.
\newblock \showarticletitle{Roberta: A robustly optimized bert pretraining
  approach}.
\newblock \bibinfo{journal}{\emph{arXiv preprint arXiv:1907.11692}}
  (\bibinfo{year}{2019}).
\newblock


\bibitem[Liu et~al\mbox{.}(2023)]%
        {liu2023fill}
\bibfield{author}{\bibinfo{person}{Zhe Liu}, \bibinfo{person}{Chunyang Chen},
  \bibinfo{person}{Junjie Wang}, \bibinfo{person}{Xing Che},
  \bibinfo{person}{Yuekai Huang}, \bibinfo{person}{Jun Hu}, {and}
  \bibinfo{person}{Qing Wang}.} \bibinfo{year}{2023}\natexlab{}.
\newblock \showarticletitle{Fill in the blank: Context-aware automated text
  input generation for mobile gui testing}. In \bibinfo{booktitle}{\emph{2023
  IEEE/ACM 45th International Conference on Software Engineering (ICSE)}}.
  IEEE, \bibinfo{pages}{1355--1367}.
\newblock


\bibitem[Liu et~al\mbox{.}(2022)]%
        {liu2022guided}
\bibfield{author}{\bibinfo{person}{Zhe Liu}, \bibinfo{person}{Chunyang Chen},
  \bibinfo{person}{Junjie Wang}, \bibinfo{person}{Yuekai Huang},
  \bibinfo{person}{Jun Hu}, {and} \bibinfo{person}{Qing Wang}.}
  \bibinfo{year}{2022}\natexlab{}.
\newblock \showarticletitle{Guided bug crush: Assist manual gui testing of
  android apps via hint moves}. In \bibinfo{booktitle}{\emph{Proceedings of the
  2022 CHI Conference on Human Factors in Computing Systems}}.
  \bibinfo{pages}{1--14}.
\newblock


\bibitem[Machiry et~al\mbox{.}(2013)]%
        {machiry2013dynodroid}
\bibfield{author}{\bibinfo{person}{Aravind Machiry}, \bibinfo{person}{Rohan
  Tahiliani}, {and} \bibinfo{person}{Mayur Naik}.}
  \bibinfo{year}{2013}\natexlab{}.
\newblock \showarticletitle{Dynodroid: An input generation system for android
  apps}. In \bibinfo{booktitle}{\emph{Proceedings of the 2013 9th Joint Meeting
  on Foundations of Software Engineering}}. \bibinfo{pages}{224--234}.
\newblock


\bibitem[Mao et~al\mbox{.}(2016)]%
        {mao2016sapienz}
\bibfield{author}{\bibinfo{person}{Ke Mao}, \bibinfo{person}{Mark Harman},
  {and} \bibinfo{person}{Yue Jia}.} \bibinfo{year}{2016}\natexlab{}.
\newblock \showarticletitle{Sapienz: Multi-objective automated testing for
  android applications}. In \bibinfo{booktitle}{\emph{Proceedings of the 25th
  International Symposium on Software Testing and Analysis}}.
  \bibinfo{pages}{94--105}.
\newblock


\bibitem[Moran et~al\mbox{.}(2018)]%
        {moran2018machine}
\bibfield{author}{\bibinfo{person}{Kevin Moran}, \bibinfo{person}{Carlos
  Bernal-C{\'a}rdenas}, \bibinfo{person}{Michael Curcio},
  \bibinfo{person}{Richard Bonett}, {and} \bibinfo{person}{Denys Poshyvanyk}.}
  \bibinfo{year}{2018}\natexlab{}.
\newblock \showarticletitle{Machine learning-based prototyping of graphical
  user interfaces for mobile apps}.
\newblock \bibinfo{journal}{\emph{IEEE Transactions on Software Engineering}}
  \bibinfo{volume}{46}, \bibinfo{number}{2} (\bibinfo{year}{2018}),
  \bibinfo{pages}{196--221}.
\newblock


\bibitem[Mozaffari et~al\mbox{.}(2022)]%
        {mozaffari2022ganspiration}
\bibfield{author}{\bibinfo{person}{Mohammad~Amin Mozaffari},
  \bibinfo{person}{Xinyuan Zhang}, \bibinfo{person}{Jinghui Cheng}, {and}
  \bibinfo{person}{Jin~LC Guo}.} \bibinfo{year}{2022}\natexlab{}.
\newblock \showarticletitle{GANSpiration: Balancing Targeted and Serendipitous
  Inspiration in User Interface Design with Style-Based Generative Adversarial
  Network}. In \bibinfo{booktitle}{\emph{Proceedings of the 2022 CHI Conference
  on Human Factors in Computing Systems}}. \bibinfo{pages}{1--15}.
\newblock


\bibitem[Najork and Wiener(2001)]%
        {najork2001breadth}
\bibfield{author}{\bibinfo{person}{Marc Najork} {and} \bibinfo{person}{Janet~L
  Wiener}.} \bibinfo{year}{2001}\natexlab{}.
\newblock \showarticletitle{Breadth-first crawling yields high-quality pages}.
  In \bibinfo{booktitle}{\emph{Proceedings of the 10th international conference
  on World Wide Web}}. \bibinfo{pages}{114--118}.
\newblock


\bibitem[Neil(2014)]%
        {neil2014mobile}
\bibfield{author}{\bibinfo{person}{Theresa Neil}.}
  \bibinfo{year}{2014}\natexlab{}.
\newblock \bibinfo{booktitle}{\emph{Mobile design pattern gallery: UI patterns
  for smartphone apps}}.
\newblock \bibinfo{publisher}{" O'Reilly Media, Inc."}.
\newblock


\bibitem[Norman(2013)]%
        {norman2013design}
\bibfield{author}{\bibinfo{person}{Don Norman}.}
  \bibinfo{year}{2013}\natexlab{}.
\newblock \bibinfo{booktitle}{\emph{The design of everyday things: Revised and
  expanded edition}}.
\newblock \bibinfo{publisher}{Basic books}.
\newblock


\bibitem[Raffel et~al\mbox{.}(2020)]%
        {raffel2020exploring}
\bibfield{author}{\bibinfo{person}{Colin Raffel}, \bibinfo{person}{Noam
  Shazeer}, \bibinfo{person}{Adam Roberts}, \bibinfo{person}{Katherine Lee},
  \bibinfo{person}{Sharan Narang}, \bibinfo{person}{Michael Matena},
  \bibinfo{person}{Yanqi Zhou}, \bibinfo{person}{Wei Li}, {and}
  \bibinfo{person}{Peter~J Liu}.} \bibinfo{year}{2020}\natexlab{}.
\newblock \showarticletitle{Exploring the limits of transfer learning with a
  unified text-to-text transformer}.
\newblock \bibinfo{journal}{\emph{The Journal of Machine Learning Research}}
  \bibinfo{volume}{21}, \bibinfo{number}{1} (\bibinfo{year}{2020}),
  \bibinfo{pages}{5485--5551}.
\newblock


\bibitem[Ren et~al\mbox{.}(2015)]%
        {ren2015faster}
\bibfield{author}{\bibinfo{person}{Shaoqing Ren}, \bibinfo{person}{Kaiming He},
  \bibinfo{person}{Ross Girshick}, {and} \bibinfo{person}{Jian Sun}.}
  \bibinfo{year}{2015}\natexlab{}.
\newblock \showarticletitle{Faster r-cnn: Towards real-time object detection
  with region proposal networks}.
\newblock \bibinfo{journal}{\emph{Advances in neural information processing
  systems}}  \bibinfo{volume}{28} (\bibinfo{year}{2015}).
\newblock


\bibitem[Rivest(1992)]%
        {rivest1992md5}
\bibfield{author}{\bibinfo{person}{Ronald Rivest}.}
  \bibinfo{year}{1992}\natexlab{}.
\newblock \bibinfo{booktitle}{\emph{The MD5 message-digest algorithm}}.
\newblock \bibinfo{type}{{T}echnical {R}eport}.
\newblock


\bibitem[Sahami~Shirazi et~al\mbox{.}(2013)]%
        {sahami2013insights}
\bibfield{author}{\bibinfo{person}{Alireza Sahami~Shirazi},
  \bibinfo{person}{Niels Henze}, \bibinfo{person}{Albrecht Schmidt},
  \bibinfo{person}{Robin Goldberg}, \bibinfo{person}{Benjamin Schmidt}, {and}
  \bibinfo{person}{Hansj{\"o}rg Schmauder}.} \bibinfo{year}{2013}\natexlab{}.
\newblock \showarticletitle{Insights into layout patterns of mobile user
  interfaces by an automatic analysis of android apps}. In
  \bibinfo{booktitle}{\emph{Proceedings of the 5th ACM SIGCHI symposium on
  Engineering interactive computing systems}}. \bibinfo{pages}{275--284}.
\newblock


\bibitem[Spencer(2009)]%
        {spencer2009card}
\bibfield{author}{\bibinfo{person}{Donna Spencer}.}
  \bibinfo{year}{2009}\natexlab{}.
\newblock \bibinfo{booktitle}{\emph{Card sorting: Designing usable
  categories}}.
\newblock \bibinfo{publisher}{Rosenfeld Media}.
\newblock


\bibitem[Su et~al\mbox{.}(2017)]%
        {su2017guided}
\bibfield{author}{\bibinfo{person}{Ting Su}, \bibinfo{person}{Guozhu Meng},
  \bibinfo{person}{Yuting Chen}, \bibinfo{person}{Ke Wu},
  \bibinfo{person}{Weiming Yang}, \bibinfo{person}{Yao Yao},
  \bibinfo{person}{Geguang Pu}, \bibinfo{person}{Yang Liu}, {and}
  \bibinfo{person}{Zhendong Su}.} \bibinfo{year}{2017}\natexlab{}.
\newblock \showarticletitle{Guided, stochastic model-based GUI testing of
  Android apps}. In \bibinfo{booktitle}{\emph{Proceedings of the 2017 11th
  Joint Meeting on Foundations of Software Engineering}}.
  \bibinfo{pages}{245--256}.
\newblock


\bibitem[Touvron et~al\mbox{.}(2023)]%
        {touvron2023llama}
\bibfield{author}{\bibinfo{person}{Hugo Touvron}, \bibinfo{person}{Thibaut
  Lavril}, \bibinfo{person}{Gautier Izacard}, \bibinfo{person}{Xavier
  Martinet}, \bibinfo{person}{Marie-Anne Lachaux},
  \bibinfo{person}{Timoth{\'e}e Lacroix}, \bibinfo{person}{Baptiste
  Rozi{\`e}re}, \bibinfo{person}{Naman Goyal}, \bibinfo{person}{Eric Hambro},
  \bibinfo{person}{Faisal Azhar}, {et~al\mbox{.}}}
  \bibinfo{year}{2023}\natexlab{}.
\newblock \showarticletitle{Llama: Open and efficient foundation language
  models}.
\newblock \bibinfo{journal}{\emph{arXiv preprint arXiv:2302.13971}}
  (\bibinfo{year}{2023}).
\newblock


\bibitem[Wang et~al\mbox{.}(2023)]%
        {wang2023enabling}
\bibfield{author}{\bibinfo{person}{Bryan Wang}, \bibinfo{person}{Gang Li},
  {and} \bibinfo{person}{Yang Li}.} \bibinfo{year}{2023}\natexlab{}.
\newblock \showarticletitle{Enabling conversational interaction with mobile ui
  using large language models}. In \bibinfo{booktitle}{\emph{Proceedings of the
  2023 CHI Conference on Human Factors in Computing Systems}}.
  \bibinfo{pages}{1--17}.
\newblock


\bibitem[Wang et~al\mbox{.}(2021)]%
        {wang2021screen2words}
\bibfield{author}{\bibinfo{person}{Bryan Wang}, \bibinfo{person}{Gang Li},
  \bibinfo{person}{Xin Zhou}, \bibinfo{person}{Zhourong Chen},
  \bibinfo{person}{Tovi Grossman}, {and} \bibinfo{person}{Yang Li}.}
  \bibinfo{year}{2021}\natexlab{}.
\newblock \showarticletitle{Screen2words: Automatic mobile UI summarization
  with multimodal learning}. In \bibinfo{booktitle}{\emph{The 34th Annual ACM
  Symposium on User Interface Software and Technology}}.
  \bibinfo{pages}{498--510}.
\newblock


\bibitem[Wu et~al\mbox{.}(2023)]%
        {wu2023webui}
\bibfield{author}{\bibinfo{person}{Jason Wu}, \bibinfo{person}{Siyan Wang},
  \bibinfo{person}{Siman Shen}, \bibinfo{person}{Yi-Hao Peng},
  \bibinfo{person}{Jeffrey Nichols}, {and} \bibinfo{person}{Jeffrey~P Bigham}.}
  \bibinfo{year}{2023}\natexlab{}.
\newblock \showarticletitle{WebUI: A Dataset for Enhancing Visual UI
  Understanding with Web Semantics}. In \bibinfo{booktitle}{\emph{Proceedings
  of the 2023 CHI Conference on Human Factors in Computing Systems}}.
  \bibinfo{pages}{1--14}.
\newblock


\bibitem[Xie et~al\mbox{.}(2020)]%
        {xie2020uied}
\bibfield{author}{\bibinfo{person}{Mulong Xie}, \bibinfo{person}{Sidong Feng},
  \bibinfo{person}{Zhenchang Xing}, \bibinfo{person}{Jieshan Chen}, {and}
  \bibinfo{person}{Chunyang Chen}.} \bibinfo{year}{2020}\natexlab{}.
\newblock \showarticletitle{UIED: a hybrid tool for GUI element detection}. In
  \bibinfo{booktitle}{\emph{Proceedings of the 28th ACM Joint Meeting on
  European Software Engineering Conference and Symposium on the Foundations of
  Software Engineering}}. \bibinfo{pages}{1655--1659}.
\newblock


\bibitem[Xie et~al\mbox{.}(2022)]%
        {xie2022psychologically}
\bibfield{author}{\bibinfo{person}{Mulong Xie}, \bibinfo{person}{Zhenchang
  Xing}, \bibinfo{person}{Sidong Feng}, \bibinfo{person}{Xiwei Xu},
  \bibinfo{person}{Liming Zhu}, {and} \bibinfo{person}{Chunyang Chen}.}
  \bibinfo{year}{2022}\natexlab{}.
\newblock \showarticletitle{Psychologically-inspired, unsupervised inference of
  perceptual groups of GUI widgets from GUI images}. In
  \bibinfo{booktitle}{\emph{Proceedings of the 30th ACM Joint European Software
  Engineering Conference and Symposium on the Foundations of Software
  Engineering}}. \bibinfo{pages}{332--343}.
\newblock


\bibitem[Yeh et~al\mbox{.}(2009)]%
        {yeh2009sikuli}
\bibfield{author}{\bibinfo{person}{Tom Yeh}, \bibinfo{person}{Tsung-Hsiang
  Chang}, {and} \bibinfo{person}{Robert~C Miller}.}
  \bibinfo{year}{2009}\natexlab{}.
\newblock \showarticletitle{Sikuli: using GUI screenshots for search and
  automation}. In \bibinfo{booktitle}{\emph{Proceedings of the 22nd annual ACM
  symposium on User interface software and technology}}.
  \bibinfo{pages}{183--192}.
\newblock


\bibitem[Zhang et~al\mbox{.}(2021)]%
        {zhang2021screen}
\bibfield{author}{\bibinfo{person}{Xiaoyi Zhang}, \bibinfo{person}{Lilian de
  Greef}, \bibinfo{person}{Amanda Swearngin}, \bibinfo{person}{Samuel White},
  \bibinfo{person}{Kyle Murray}, \bibinfo{person}{Lisa Yu}, \bibinfo{person}{Qi
  Shan}, \bibinfo{person}{Jeffrey Nichols}, \bibinfo{person}{Jason Wu},
  \bibinfo{person}{Chris Fleizach}, {et~al\mbox{.}}}
  \bibinfo{year}{2021}\natexlab{}.
\newblock \showarticletitle{Screen recognition: Creating accessibility metadata
  for mobile applications from pixels}. In
  \bibinfo{booktitle}{\emph{Proceedings of the 2021 CHI Conference on Human
  Factors in Computing Systems}}. \bibinfo{pages}{1--15}.
\newblock


\bibitem[Zhou and Li(2021)]%
        {zhou2021large}
\bibfield{author}{\bibinfo{person}{Xin Zhou} {and} \bibinfo{person}{Yang Li}.}
  \bibinfo{year}{2021}\natexlab{}.
\newblock \showarticletitle{Large-scale modeling of mobile user click behaviors
  using deep learning}. In \bibinfo{booktitle}{\emph{Proceedings of the 15th
  ACM Conference on Recommender Systems}}. \bibinfo{pages}{473--483}.
\newblock


\end{thebibliography}

\appendix
\section{Appendix}


\subsection{Web Interface for UI Validation}
\begin{figure*}[htp]
	\centering
	\includegraphics[width = 0.75\linewidth]{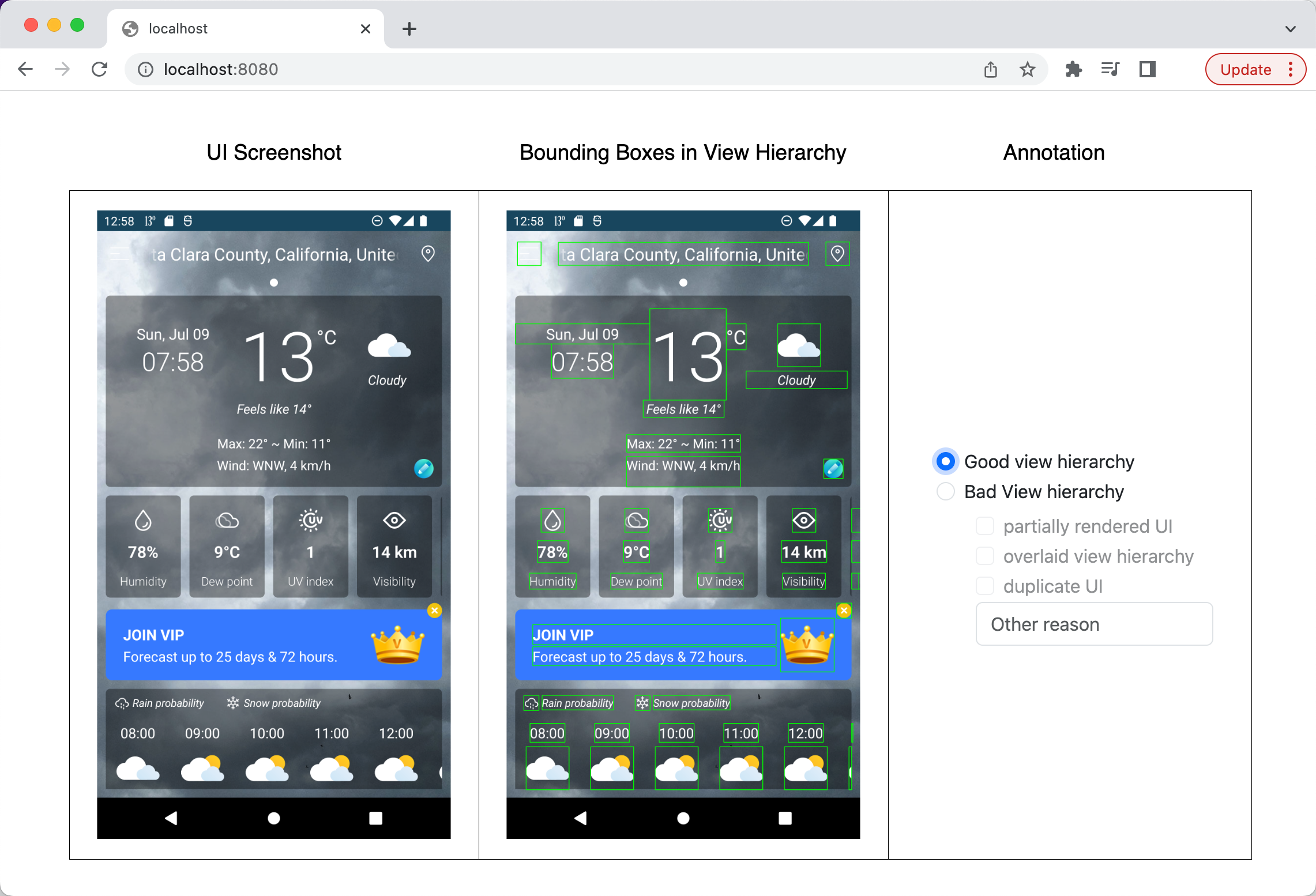}
	\caption{In the interface, we highlight the UI screenshot along with the bounding boxes extracted from the view hierarchy. Annotators can flag the UI as invalid and select the reasons, such as partially rendered UI, overlaid view hierarchy, duplicate UI, or explain the other reason.}
	\label{fig:interface}
\end{figure*}

\end{document}